\documentclass[twocolumn]{aastex631}
\usepackage{epsfig,psfig}
\usepackage{amsmath}
\usepackage{graphics}
\usepackage{longtable}
\usepackage{geometry}

\newcommand{\E}[1]{\times 10^{#1}}
\newcommand{\s}{\,{\rm s}}      \newcommand{\ps}{\,{\rm s}$^{-1}$}
    \newcommand{\Msun}{{\rm M}_{\rm \odot}}
\newcommand{\cm}{\,{\rm cm}}    \newcommand{\km}{\,{\rm km}}
\newcommand{\kpc}{\,{\rm kpc}}
        \newcommand{\K}{\,{\rm K}}

\newcommand{\degree}{$^\circ$}
\newcommand{\HI}{H~{\sc i}}
\newcommand{\HII}{H~{\sc ii}}
\newcommand{\vlsr}{$V_{\rm LSR}$}       \newcommand{\tmb}{$T_{\rm mb}$}
\newcommand{\twCO}{$^{12}$CO}  \newcommand{\thCO}{$^{13}$CO}
\newcommand{\CeiO}{C$^{18}$O}
\newcommand{\myemail}{xinzhou@pmo.ac.cn}
\shorttitle{Molecular gas distribution toward the inner\&outer Galaxy}
\shortauthors{Zhou et al.}

\begin{document}

\title{Molecular Gas Distribution toward the Inner and Outer Galaxy Revealed by MWISP\\
--- the Galactic Longitude 45\degree--60\degree and 120\degree--130\degree
}

\email{\myemail}

\author{Xin Zhou}
\author{Ji Yang}
\author{Yan Sun}
\author{Qing-Zeng Yan}
\author{Lixia Yuan}
\author{Yang Su}
\author{Xuepeng Chen}
\author{Shaobo Zhang}

\affil{Purple Mountain Observatory and Key Laboratory of Radio Astronomy, Chinese Academy of Sciences, 10 Yuanhua Road, Nanjing 210033, People's Republic of China; \myemail}

 
 

\begin{abstract}
Molecular clouds (MCs) are cradles of star and planet formation, thereby playing an important role in the evolution of galaxies. 
Based on the unbiased Milky Way Imaging Scroll Painting (MWISP) survey data of \twCO, \thCO, and \CeiO\ (J=1--0) line emission in two regions toward the inner and outer Galaxy, i.e.\ the G50 (44.75\degree$\le l \le60.25$\degree) and G120 (119.75\degree$\le l \le130.25$\degree) regions, the distribution of molecular gas is studied.
Both regions have Galactic latitudes of $|b|\le5.25$\degree.
A catalog containing 24724 MCs is constructed from the data.
In our proximity, several molecular structures with large angular scales and small velocity dispersions are discovered, resembling curtains of mist.
Beyond the nearby molecular gas, a clear aggregation of MCs along coherent structures in the Galactic plane is visible, sketching spiral arm structures.
Nevertheless, the aggregation of MCs is also detected in the inter-arm region between the Perseus and Outer arms in the G50 region. 
The Galactic molecular disk in this inter-arm region is found to be thinner than that in the adjacent spiral arm region.
In addition, the thickness of the Galactic molecular disk examined here is found to be correlated with the warp of it, indicating their homologous origins. The molecular disk has a typical thickness of $\sim$220~pc in the inner Galaxy.
Moreover, the dispersion of the MC systemic velocity decreases with increasing galactocentric radius, resulting in lower kinematic distance uncertainties at larger radii.
However, the Perseus arm segment in the G120 region exhibits a relatively large cloud-to-cloud velocity dispersion and split components in its MC velocity distribution. 
\end{abstract}



\section{Introduction} \label{sec:intro}
Molecular clouds (MCs), as cradles of star and planet formation, play a crucial role in the evolution of galaxies. 
The study of MCs is essential for understanding the physical processes that govern star formation in galaxies \citep[e.g.,][]{BerginTafalla2007, Chevance+2020}.
MCs are mainly composed of molecular hydrogen (H$_2$), along with other molecules such as CO, NH$_3$, etc. 
The presence of these molecules facilitates the cooling of the gas \citep[especially CO;][]{GoldsmithLanger1978} and its subsequent gravitional collapse, leading to the formation of dense cores that eventually evolve into stars and planetary systems.
While the overall evolutionary path appears clear, MCs exhibit significant diversity in their morphology and physical state, suggesting different origins and fates. Furthermore, the star-forming process drives a strong feedback on the parent MC, making the overall evolution nonlinear. 

Molecular spectral lines can provide a wealth of information about the physical state of MCs. Nevertheless, the most abundant molecule, H$_2$, is difficult to detect directly due to its lack of a permanent dipole moment. Consequently, carbon monoxide (CO) is often used as a tracer for molecular gas, such as \twCO\ (J=1--0) line, which can be used to trace very diffuse molecular gas. 
There are numerous large Galactic plane surveys of these CO lines, e.g., the Columbia CO survey \citep{Dame+2001, DameThaddeus2022}, the Galactic Ring Survey \citep[GRS][]{Jackson+2006}, the FOREST Unbiased Galactic plane Imaging survey \citep[FUGIN][]{Umemoto+2017}, the Mopra CO survey \citep{Burton+2013, Braiding+2015, Braiding+2018, Cubuk+2023}, etc.
In addition to CO, there are other various molecular lines that can be used to study molecular gas in different physical states, especially dense gas deep in clouds.

Based on large-scale CO surveys \citep[e.g.,][]{Sanders+1986, Clemens+1986, Heyer+1998, Dame+2001, Jackson+2006, Schuller+2017, Schuller+2021}, the distribution of molecular gas and the physical properties of MCs have been studied \citep[e.g.,][]{Solomon+1987, Heyer+2001, Rice+2016, Rathborne+2009, Roman-Duval+2010, Duarte-Cabral+2021, Neralwar+2022b}.
The molecular gas in the Milky Way is mainly distributed along spiral arms \citep[as shown by][etc.]{Dame+2001, Schuller+2021, Sun+2024}.
An MC can span hundreds of parsecs, while its cores are only sub-parsec in size. Cloud structures exhibit a wide range of sizes and shapes. 
Some MCs are filamentary in shape, while others are globular, ring-like, or irregular \citep[e.g.,][and references therein]{Yuan+2021, Neralwar+2022a, Clarke+2022}.
These structures form at different scales within the interstellar medium (ISM) through gravitational collapse and turbulence, as well as under the influence of magnetic fields and stellar feedback.
MCs are in a transition zone between galaxy-scale structures and dynamics (kilo-parsecs), and the dominance of local processes (e.g., parsec- or sub-parsec-scale stellar feedback).
Consequently, MC evolution is complex, yet provides insight into both the large-scale features of the galaxy, small-scale star formation processes, and their interconnections.

The primary objective of this study is to investigate the distribution of MCs in the Galactic disk.
The unbiased survey data of \twCO, \thCO, and \CeiO\ (J=1--0) line emission, i.e.\ the Milky Way Imaging Scroll Painting (MWISP) survey, is used. 
The observations and data quality are described in Section~\ref{sec:obs}. The results are presented in Section~\ref{sec:result}, including several parts: the overview of the molecular gas distribution is presented in Section~\ref{sec:over}, the MC identification techniques are presented in Section~\ref{sec:mcident}, the location of MCs in the Milky Way is presented in Section~\ref{sec:mcloc}, and the inferred structure of the Galactic disk is presented in Section~\ref{sec:disk}.
Section~\ref{sec:sum} summarizes our conclusions.

\section{Observations} \label{sec:obs}
\begin{figure*}[ptbh!]
\centerline{{\hfil\hfil
\psfig{figure=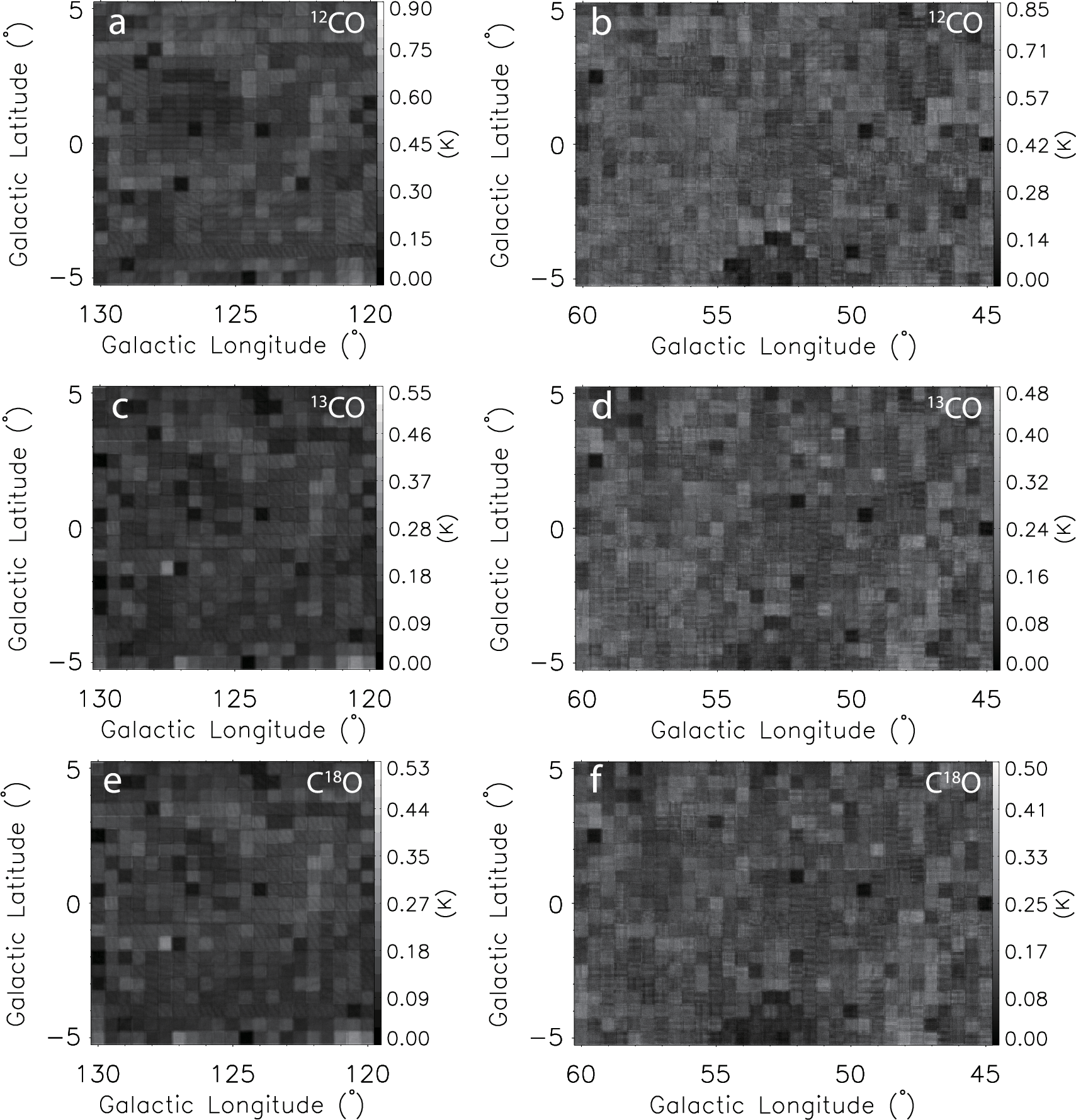,height=5.5in,angle=0, clip=}
\hfil\hfil}}
\caption{RMS noise maps of \twCO\ (J=1--0), \thCO\ (J=1--0), and \CeiO\ (J=1--0) line emission.}
\label{f:rmsmap}
\end{figure*}
\begin{figure*}[ptbh!]
\centerline{{\hfil\hfil
\psfig{figure=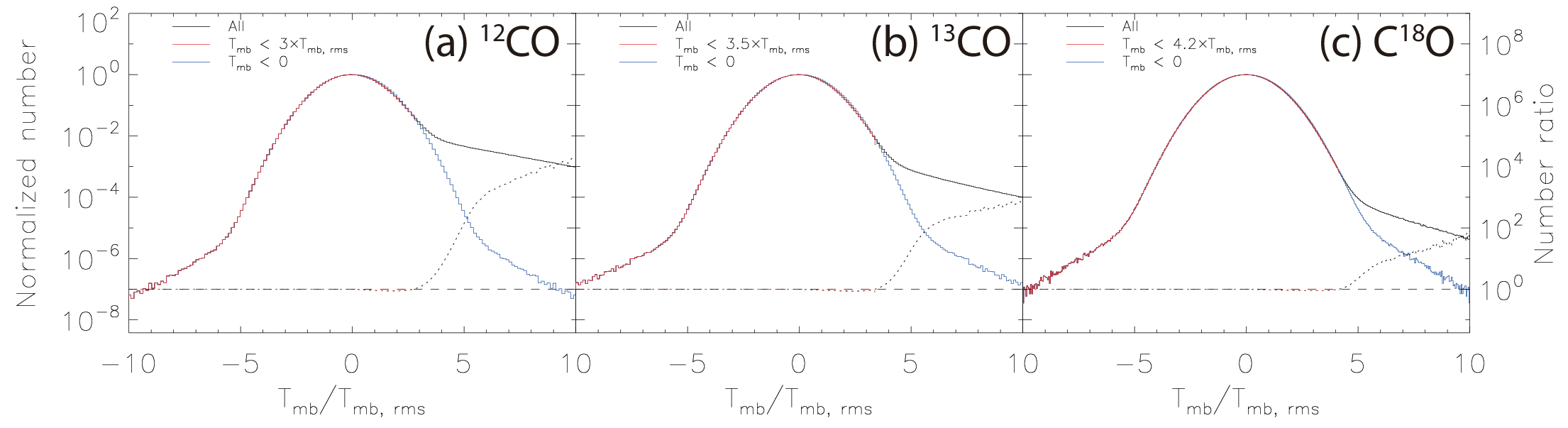,width=6in,angle=0, clip=}
\hfil\hfil}}
\caption{Signal-to-noise ratio distribution of \twCO\ (J=1--0), \thCO\ (J=1--0), and \CeiO\ (J=1--0) line emission. There are approximately $5.17\E{9}$ voxels in total. Normalized numbers of all voxels are shown by black solid lines, and selected voxels with signal-to-noise ratios below 3 for \twCO, 3.5 for \thCO, and 4.2 for \CeiO\ are shown by red solid lines. The negative intensity voxels are symmetrically mirrored to positive values via the zero point (blue). Number ratios between all voxels and the negative plus mirror voxels are shown by black dotted lines, and ratios for the selected voxels are shown by red dotted lines. 
}
\label{f:snratio}
\end{figure*}

The CO line emission was observed using the Purple Mountain Observatory (PMO) 13.7~m millimeter-wavelength telescope located in Delingha, China, as part of the MWISP project\footnote{http://english.dlh.pmo.cas.cn/ic/} \citep[see][and references therein, for details]{Su+2019,Sun+2020}. 
We mapped two regions toward the inner and outer Galaxy, namely the G50 ($l=44$.75\degree--60.25\degree) and G120 ($l=119$.75\degree--130.25\degree) regions, both with the Galactic latitude $|b|\le5.25$\degree.
The \twCO~(J=1--0), \thCO~(J=1--0), and \CeiO~(J=1--0) lines were observed simultaneously by the $3\times3$ multibeam sideband separation receiver, i.e.\ the Superconducting Spectroscopic Array Receiver \citep[SSAR,][]{Shan+2012}. 
The total error in pointing and tracking was within $5''$, and the half-power beam width (HPBW) was $\sim51''$.
The spectral resolutions of the three CO lines were 0.17~\km\ps\ for \twCO~(J=1--0) and 0.16~\km\ps\ for both \thCO~(J=1--0) and \CeiO~(J=1--0).
Following the standard procedure of the MWISP CO line survey, the data were meshed with a grid spacing of $30''$.
A linear fit was performed in the baseline subtraction. 
Data with bad baselines were excluded, however, minor RMS noise inhomogeneities were introduced. 
RMS noise maps are shown in Figure~\ref{f:rmsmap}.
Typical RMS levels are $\sim$0.45~K per channel for \twCO~(J=1--0) and $\sim$0.24~K per channel for \thCO~(J=1--0) and \CeiO~(J=1--0).
As demonstrated in Figure~\ref{f:snratio}, the signal-to-noise ratio distributions indicate that the majority of signals are separate from noise at the signal-to-noise ratio of $\sim$3 for \twCO\ (J=1--0) and \thCO\ (J=1--0) and the ratio of $\sim$4 for \CeiO\ (J=1--0).
All data were processed using dedicated pipelines by the MWISP working group and the GILDAS/CLASS package\footnote{http://www.iram.fr/IRAMFR/GILDAS}.

\section{Results} \label{sec:result}
\subsection{Overview} \label{sec:over}
\begin{figure*}[ptbh!]
\centerline{{\hfil\hfil
\psfig{figure=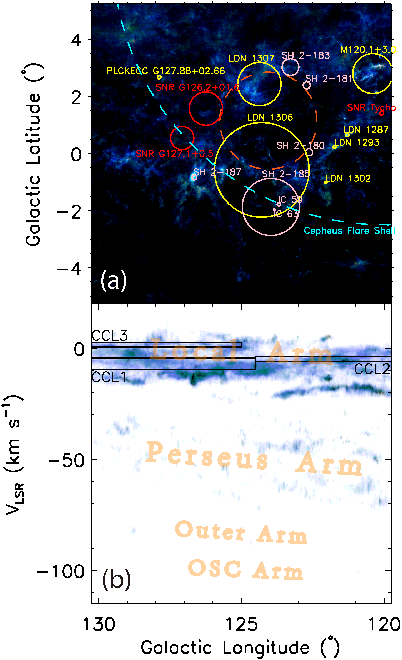,width=4in,angle=0, clip=}
\hfil\hfil}}
\caption{Pseudo-tricolor image of the integrated intensity of \twCO\ (J=1--0) (blue), \thCO\ (J=1--0) (green), and \CeiO\ (J=1--0) (red) line emission in the velocity range of $-$115 to 20 \km\ps\ ({\it top}), and Galactic longitude-velocity ({\it l-v}) map of the three lines for the same field ({\it bottom}), for the G120 region.
The data were initially moment-masked \citep[see][for reference]{Dame2011} to suppress noise while integrating over the large velocity ranges, thus the minimum values of the maps are zero. The \CeiO\ data was also masked using the \thCO\ mask. The intensity is in linear scale, with the maximum values of 57.3, 17.3, and 3.0 K~\km\ps\ for \twCO, \thCO, and \CeiO, respectively. 
Integrated intensities greater than these maxima are truncated for better visibility.
Some well-studied objects in the region are indicated by circles, including clouds \citep[yellow;][]{Lynds1962,Yang+1990,Planck+2016}, \HII\ regions \citep[pink;][]{Dubout-Crillon1976, Karr+2005}, and supernova remnants \citep[red;][and references therein]{Zhou+2023}. 
Most clouds were initially identified as dark nebulae in the optical band by \cite{Lynds1962}, while others were discovered by CO emission \citep[e.g., M120.1$+$3.0;][]{Yang+1990} and sub-mm observations \citep[e.g., PLCKECC G127.88$+$02.66;][]{Planck+2016}. 
The cyan dashed line shows the approximate extent of the Cepheus Flare Shell, which contains different MC complexes \citep{Grenier+1989, Kirk+2009}.
A prominent shell-like MC is indicated by an orange dashed circle, which is located around the Galactic coordinate (124\degree, 1\degree) and extends $\sim$3\degree\ \citep{Chenxp+2023}.
In the {\it l-v} map, the approximate locations of the spiral arms are indicated, i.e.\ the Local, Perseus, Outer, and Outer Scutum-Centaurus (OSC) arms \citep[e.g.,][]{Vallee2008, Sun+2024}. The CO emission detected in the Outer and OSC arms is weak.
Regions CCL1, CCL2, and CCL3 indicate MCs with large angular sizes and narrow line widths.
The velocity ranges of these regions are determined by the velocity dispersion of the corresponding MCs.
}
\label{f:g120intlv}
\end{figure*}

\begin{figure*}[ptbh!]
\centerline{{\hfil\hfil
\psfig{figure=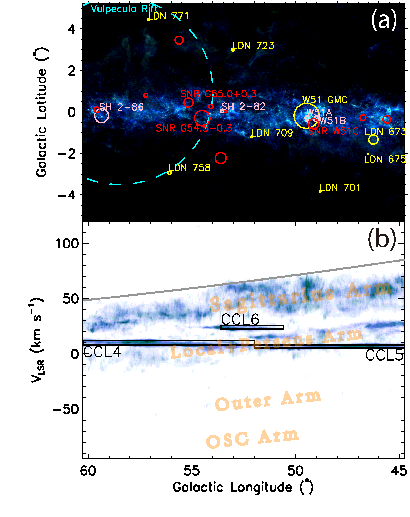,width=6in,angle=0, clip=}
\hfil\hfil}}
\caption{
Same as Figure~\ref{f:g120intlv}, but for the G50 region.
The velocity range of the integrated intensity map is $-$95 to 120 \km\ps, and the corresponding maximum values are 101.3, 31.4, and 9.9 K~\km\ps.
Some well-studied objects in the region are indicated as well, including those in the W51 region. The W51 region contains different kinds of objects and is an active star-forming region \citep[e.g.,][and references therein]{Carpenter+1998}.
The cyan dashed line shows the approximate extent of the Vulpecula Rift \citep{DutraBica2002}, which contains different MC complexes.
In the {\it l-v} map, MCs with large angular sizes and narrow line widths are indicated by regions CCL4, CCL5, and CCL6. 
The approximate locations of the Sagittarius, Local+Perseus, Outer, and OSC spiral arms are indicated \citep[e.g.,][]{Vallee2008, Sun+2024}.
A grey solid line shows the position of the tangent points with increased velocity by 15~\km\ps, estimated based on the Galactic rotation curve (refer to Section~\ref{sec:mcloc} for details).
This additional increase in velocity accounts for the peculiar velocity of MCs. 
After increasing the velocity by 15~\km\ps, the tangent points follow the terminal velocity of the \twCO\ (J=1--0) emission well.
}
\label{f:g50intlv}
\end{figure*}

The overall spatial distribution of the CO emission along with some well-studied objects in the regions is illustrated in Figures~\ref{f:g120intlv} and \ref{f:g50intlv}.
Toward the inner Milky Way, in the G50 region, the Galactic plane is clearly visible. MCs, \HII\ regions, and supernova remnants (SNRs) are mainly distributed along the Galactic plane. There are also plenty of molecular gas detected at high Galactic latitude. Some high latitude gas is within the Cepheus Flare Shell and the Vulpecula Rift, in the G120 and G50 regions, respectively, which are located nearby. The Cepheus Flare Shell and the Vulpecula Rift are expansive, sparsely populated dark nebula regions \citep{Hubble1934,DutraBica2002}. These regions contain MC complexes located at various distances, e.g., from $\sim$160 to $\sim$800~pc for the Cepheus Flare Shell \citep{Kirk+2009}.
Shell-like and filamentary structures are widely distributed in both the G120 and G50 regions, e.g., a prominent shell-like MC located around the Galactic coordinates (124\degree, 1\degree) and extends $\sim$3\degree\ \citep[see the top panel of Figure~\ref{f:g120intlv}; refer to][for further details]{Chenxp+2023} and the filamentary MC M120.1$+$3.0 \citep{Yang+1990, SunL+2023}.
We note that some previously studied MCs are focused only on their central regions.

The Galactic longitude-velocity (l-v) maps in Figures~\ref{f:g120intlv} and \ref{f:g50intlv} provide us many information on molecular gas distribution at both large and small scales. 
The tendency for molecular gas to concentrate in coherent structures in l-v maps is evident, sketching spiral arm structures \citep[e.g.,][]{Dame+2001, DameThaddeus2011}.
Strong CO emission is detected in the vicinity of the Sun, with $V_{\rm LSR}\sim0$~\km\ps, potentially associated with the Local arm. 
Molecular gas with negative velocities in the G50 region is located beyond the solar circle at a distance greater than $\sim$10~kpc, resulting in a weak emission signal.
Weak \twCO\ emission is also detected on the Outer Scutum-Centaurus (OSC) arm, for instance at $V_{\rm LSR}\sim-100$~\km\ps\ in the G120 region \citep{Sun+2015, Sun+2024}. 
In the G50 region, the Galactic rotation curve confines the upper limit of the MC velocity at different Galactic longitudes, i.e.\ the tangent points. 
However, molecular gas with velocities greater than those of the adjacent tangent point MCs is also detected, e.g., at $V_{\rm LSR}\sim65$~\km\ps\ and $l\sim53$\degree, although this could be simply due to peculiar motions of the order of 15~\km\ps\ (see Section~\ref{sec:mcloc} for more details).
Additionally, numerous small vertical strips are also observed in the l-v maps. These features can be driven by star-formation activity \citep[e.g., LDN 1287;][]{Yang+1991, Yang+1995} or SNRs \citep{Zhou+2023}, etc. 

\begin{figure*}[ptbh!]
\centerline{{\hfil\hfil
\psfig{figure=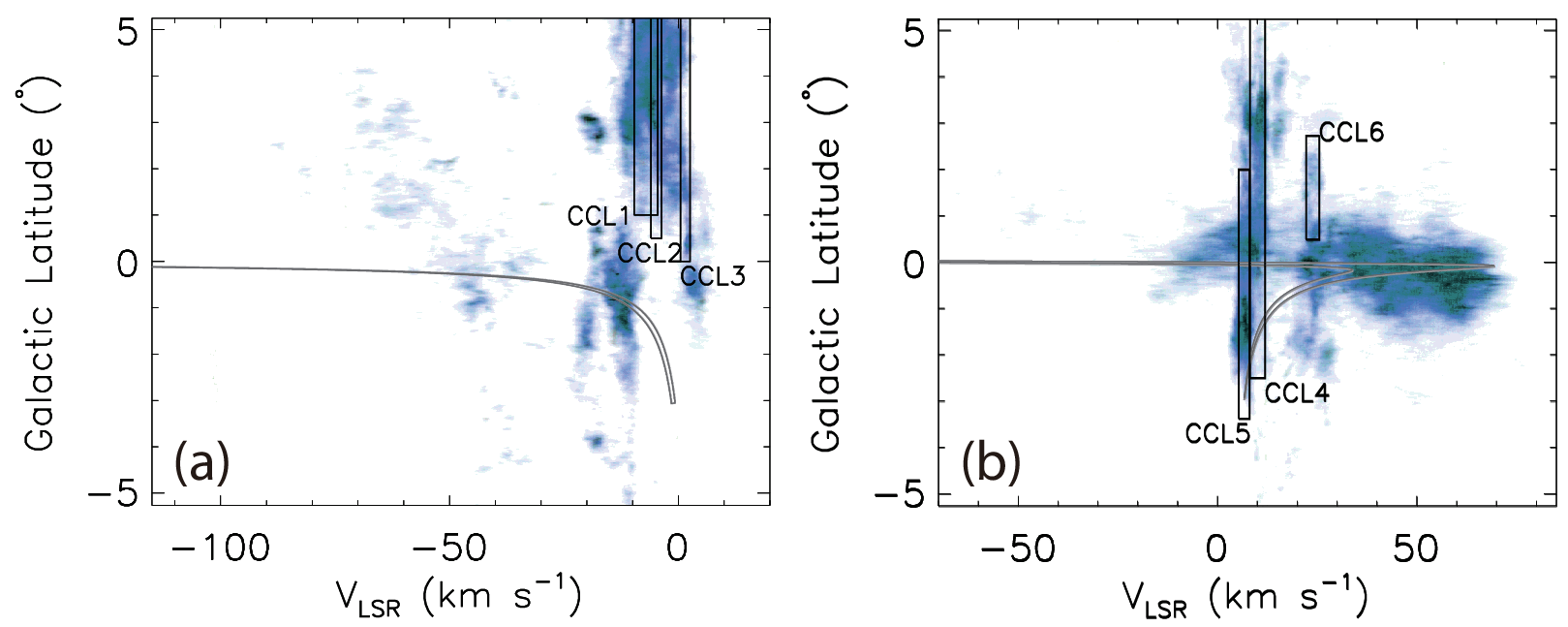,width=6in,angle=0, clip=}
\hfil\hfil}}
\caption{Galactic velocity-latitude (v-b) maps of \twCO\ (J=1--0) (blue), \thCO\ (J=1--0) (green), and \CeiO\ (J=1--0) (red) line emission in the Galactic longitude ranges of 119.75\degree to 130.25\degree ({\it left}) and 44.75\degree to 60.25\degree ({\it right}). The data were initially moment-masked to suppress noise.
The \CeiO\ data was also masked using the \thCO\ mask.
Regions of the same MCs as indicated in the l-v maps in Figures~\ref{f:g120intlv} and \ref{f:g50intlv} are shown. The velocity ranges of the regions are determined by the corresponding inter-cloud velocity dispersion.
Grey solid lines indicate the position of the Galactic midplane corresponding to the Galactic longitude ranges, estimated based on the Galactic rotation curve (refer to Section~\ref{sec:mcloc} for details). The midplane closer than 300~pc is not shown.
}
\label{f:vbmap}
\end{figure*}
Galactic velocity-latitude (v-b) maps are shown in Figure~\ref{f:vbmap}.
High latitude molecular gas and near b=0\degree\ disk gas have different distributions.
The width of the gas distribution along the Galactic latitude gives an indication of the distance.
Because of the projection effect, the angular scale height is smaller for clouds that are farther away, even if they are at the same physical scale height. 
Molecular gas is widely distributed along the Galactic latitude at close distances, while it is relatively concentrated on the Galactic disk at large distances.
The different distribution of the angular scale height can help to distinguish molecular gas with the same velocity but at different distances to some extent, i.e.\ for those with positive velocities in the G50 region.
As shown in the v-b maps (Figure~\ref{f:vbmap}), the relative position of the molecular gas with respect to the Galactic midplane varies at different velocities, i.e., at different distances. It demonstrates the warp of the Galactic disk.

\begin{figure*}[ptbh!]
\centerline{{\hfil\hfil
\psfig{figure=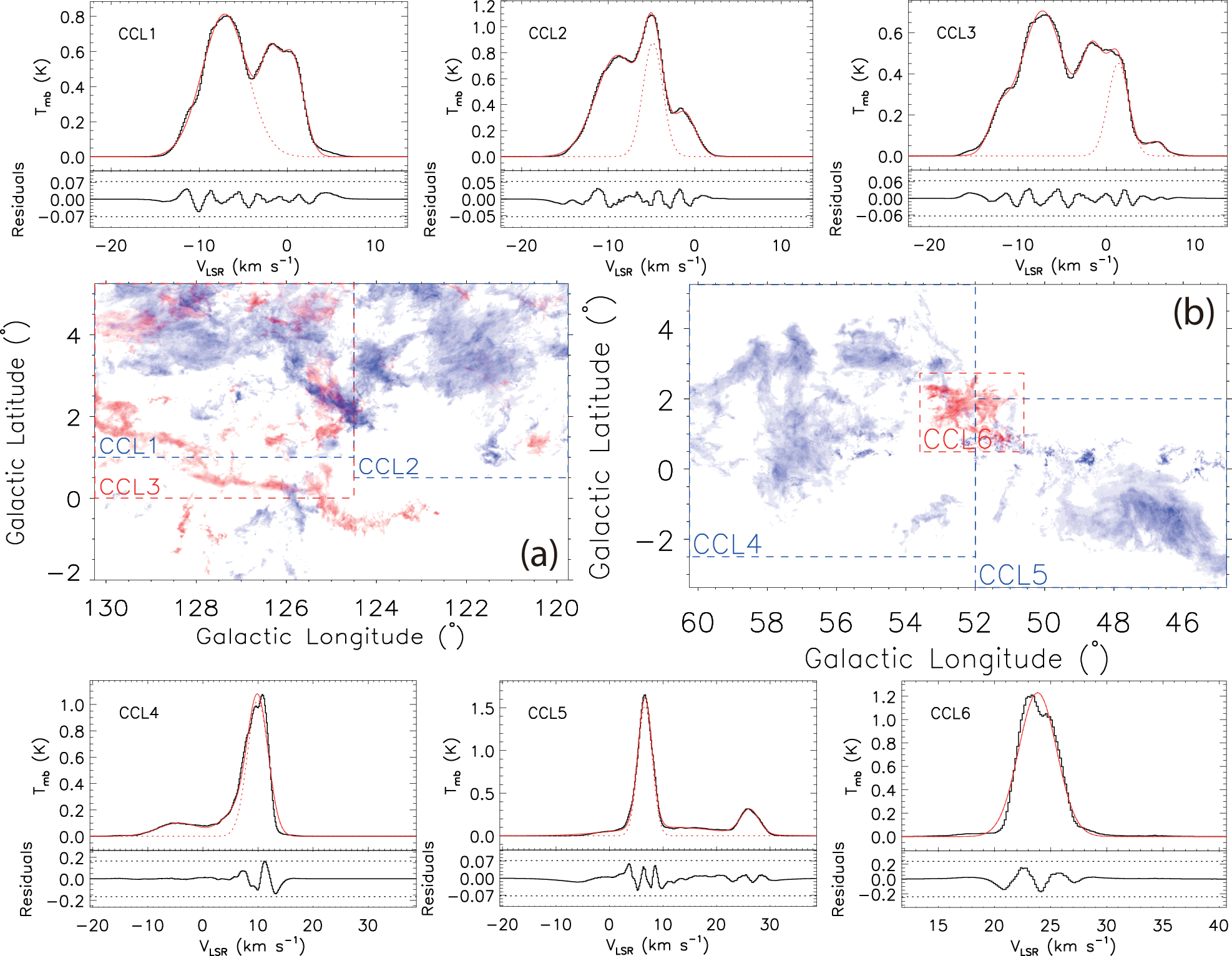,width=6in,angle=0, clip=}
\hfil\hfil}}
\caption{{\it Middle row}: \twCO\ (J=1--0) integrated intensity maps of selected MCs.
The MCs are the same as those indicated in the l-v and v-b maps (see Figures~\ref{f:g120intlv}, \ref{f:g50intlv}, and Figure~\ref{f:vbmap}). 
The velocity intervals of the maps are $-9.7$ to $-3.7$ \km\ps\ (blue) and 0.2 to 2.5 \km\ps\ (red) for the G120 region (left), and 5.3 to 11.9 \km\ps\ (blue) and 22.1 to 25.5 \km\ps\ (red) for the G50 region (right).
The velocity ranges are determined by the velocity dispersions of the corresponding MCs. CCL1 and CCL2 have similar velocities and are covered by the same velocity range. The same is true for for CCL4 and CCL5.
Different MCs are indicated by dashed rectangles, with colors corresponding to those of the integrated intensity map.
The cloud data were initially extracted from the datacube based on connectivity in position-position-velocity space (see Section~\ref{sec:mcident} for details) to suppress background emission. 
Nevertheless, there is still some background emission left, particularly in the G50 region. Bright CO emission can be seen in the velocity range of 5.3 to 11.9 \km\ps\ at $b\sim0$\degree, which originates from the Galactic disk at large distances.
Only areas with cloud cover are shown, which are large but not the entire observation coverage.
The integrated intensities in the maps are all in linear scale. 
The minimum values of the maps are zero.
The corresponding maximum values of the maps are 22.8, 11.7, 34.8, and 14.0~K~\km\ps, respectively.
Integrated intensities greater than the maximum values are truncated for better visibility.
{\it Other six pannels}: average \twCO~(J=1--0) spectra extracted from the corresponding regions of CCL1, CCL2, CCL3, CCL4, CCL5, and CCL6, respectively, together with their best-fit Gaussian models (red solid lines) and residuals. 
The $5\sigma$ levels of the residuals are shown by the black dotted lines.
The Gaussian components peak at $-7.1, -2.0$, and 0.5~\km\ps\ for the CCL1 region, $-8.9, -4.8$, and $-1.2$~\km\ps\ for the CCL2 region, $-12.2, -7.2, -1.5$, 1.3, and 5.5~\km\ps\ for the CCL3 region, $-4.3$, 5.8, and 10.0~\km\ps\ for the CCL4 region, 6.6, 11.0, and 26.1~\km\ps\ for the CCL5 region, and 23.8~\km\ps\ for the CCL6 region. The individual Gaussian components of the corresponding MCs are shown by red dotted lines. The fitted parameters of these components are presented in Table~\ref{tab:cclpar}.
}
\label{f:intmap1}
\end{figure*}

\begin{table*}
\begin{center}
\caption{Parameters of the selected MCs (see Figure~\ref{f:intmap1}).\label{tab:cclpar}}
\begin{tabular}{ccccccccccc}
\tableline\tableline
Cloud & {\it l}\tablenotemark{a} & {\it b}\tablenotemark{a} & {\it l}$_{\rm span}$ & {\it b}$_{\rm span}$ & \tmb$_{\rm, peak}$\tablenotemark{b} & \vlsr\tablenotemark{b} & $\sigma_v$\tablenotemark{b}&$N({\rm H}_2)$\tablenotemark{c}&$D$&$M$\tablenotemark{g}\\
&(\degree)&(\degree)&(\degree)&(\degree)&(K)&(\km\ps)&(\km\ps)&($10^{20} {\rm cm}^{-2}$)&(kpc)&($10^3 \Msun$)\\
\tableline
CCL1&127.38&3.13&5.75&4.25&0.8&$-7.1$&2.6& 9.4&0.7\tablenotemark{d}&61\\
CCL2&122.13 &2.88 &4.75 &4.75 &0.9 &$-4.8$ &1.1&4.5&0.6\tablenotemark{d}&17 \\
CCL3&127.38 &2.63 &5.75 &5.25 &0.5 &1.3 &1.2&2.4&0.19\tablenotemark{e}&1.3 \\
CCL4&56.13 &1.38 &8.25 &7.75 &1.0 &10.0 &1.9 & 8.8&0.6\tablenotemark{d}&57\\
CCL5&48.38 &$-0.69$ &7.25 &5.38 &1.5 &6.6 &1.3& 9.2&0.4\tablenotemark{f}&19 \\
CCL6&52.10 &1.61 &3.00 &2.24 &1.2 &23.8 &1.7 & 9.3&1.5\tablenotemark{d}&46\\
\tableline
\end{tabular}
\tablenotetext{a}{Center of the cloud region.}
\tablenotetext{b}{Peak $T_{\rm mb}$, systemic velocity, and velocity dispersion estimated by Gaussian fitting to the corresponding cloud component.}
\tablenotetext{c}{H$_{2}$ column density, estimated using the conversion factor $N({\rm H_2})/W(^{12}{\rm CO})=1.8\E{20} \cm^{-2}(\K \km \s^{-1})^{-1}$ \citep{Dame+2001}.}
\tablenotetext{d}{Near kinematic distance estimated using the Galactic rotation curve, refer to Section~\ref{sec:over} for details.}
\tablenotetext{e}{Distance derived by \cite{Chenxp+2023} based on an extinction-parallax method.}
\tablenotetext{f}{Distance derived by \cite{Su+2020} and \cite{Yan+2020} based on an extinction-parallax method.}
\tablenotetext{g}{Cloud mass, estimated as $M=N({\rm H_2}) A D^2 \mu({\rm H_2}) m_{\rm H}$, where $A$ is the angular area of the cloud and $\mu({\rm H_2})=2.8$ is the average molecular weight per hydrogen molecule \citep{Kauffmann+2008}, and $m_{\rm H}$ is the hydrogen atom mass.}
\end{center}
\end{table*}

There are several large scale strips around 0~\km\ps\ detected in both l-v (the bottom panels of Figures~\ref{f:g120intlv} and \ref{f:g50intlv}) and v-b (Figure~\ref{f:vbmap}) maps.
We separate several notable structures into six clouds: CCL1, CCL2, CCL3, CCL4, CCL5, and CCL6.
These clouds are initially extracted from the datacube based on connectivity in the position-position-velocity space to suppress background emission, nevertheless, some background emission remains (see Section~\ref{sec:mcident} for details). 
Figure~\ref{f:intmap1} shows their integrated intensity maps and average spectra. Gaussian fitting is performed on the average spectra to decompose the corresponding cloud components, and the parameters of these MCs are presented in Table~\ref{tab:cclpar}. 
The 1.3~\km\ps\ MC (CCL3) in the G120 region is more fibrous than the others, and its major filament is part of the large-scale Cassiopeia Filament studied by \cite{Chenxp+2023}. 
The 6.6~\km\ps\ MC (CCL5) in the G50 region is part of the large River cloud, which was studied by \cite{Su+2020} and \cite{Yan+2020}. 
In the G50 region, there are some bright CO emission at $b\sim0$\degree, which are background emission at large distances.
The large MCs have very small velocity widths.
Similar clouds have been also found in the \thCO\ (J=2--1) line emission of the SEDIGISM survey, which were classified as wispy clouds \citep{Schuller+2021}. They have smaller sizes and narrower line widths in the \thCO\ (J=2--1) line emission than in the \twCO\ (J=1--0) line emission detected here.

These clouds have large extensions and small velocity dispersions.
They extend beyond the boundary of the observation coverage, except for CCL6.
The MC CCL6 is extracted from the datacube in its entirety based on connectivity in the position-position-velocity space (see Section~\ref{sec:mcident} for details).
Most of these MCs have systemic velocities around 0~\km\ps\, which suggests they are close to us. 
The distances of MCs CCL3 and CCL5 are $\sim$190 and $\sim$400~pc, respectively, which are estimated based on an extinction-parallax method \citep{Su+2020, Yan+2020, Chenxp+2023}. 
The distances of other MCs are estimated using the Galactic rotation curve of \cite{BrandBlitz1993} with the Sun's galactocentric distance of 8.15~kpc and the circular rotation speed of 236~\km\ps\ \citep{Reid+2019}.
Both MCs CCL4 and CCL6 exhibit kinematic distance ambiguities. 
If placed at their far kinematic distances ($\sim$8.5~kpc), their derived sizes would exceed 0.4~kpc and 1~kpc, and their masses would be at least $1.5\E{6}~\Msun$ and $1\E{7}~\Msun$, repectively.
These extreme physical parameters suggest that the MCs are located at their near kinematic distances.
For nearby MCs (CCL1, CCL2, and CCL4), the relative error of the kinematic distance is large.
These MCs containing abundant molecular gas are probably not within the Local Bubble and therefore at a distance greater than 165~pc \citep{Zucker+2022}.
According to the Larson velocity dispersion-size relation \citep{Larson1981}, even assuming the smallest distance of 165~pc for CCL1, CCL2, and CCL4, the velocity dispersion of all the clouds should be greater than 3~\km\ps\ at least, which is significantly larger than the velocity dispersion detected (see Table~\ref{tab:cclpar}). In other words, the size of these MCs are too large compared to their narrow velocity width. 
Unlike traditional MCs that follow the Larson relation, these MCs are probably thin in the line-of-sight (LoS) direction.
This is also supported by the estimation of their density. Assuming the LoS extent of these MCs is comparable to their plane-of-sky size, their density would be less than 10~$\cm^{-3}$, which is significantly lower than the critical density of \twCO~(J=1--0) emission ($10^3~\cm^{-3}$) and even more than an order of magnitude lower than the \twCO\ effective excitation density \citep{Shirley2015}.
Considering their large scale, small thickness, and widespread distribution in different directions, these MCs seem to be sheet-like clouds, forming a curtain of mist surrounding us, which we will refer to as curtain MCs.
The curtain MCs can have different structures in different layers. 

\subsection{MC identification} \label{sec:mcident}
\begin{table*}
\tiny 
\begin{center}
\caption{Basic parameters derived by \twCO~(J=1--0) emission for all MCs.\label{tab:mcpar}}
\begin{tabular}{ccccccccccccc}
\tableline\tableline
Cloud id.\tablenotemark{a}& {\it l}\tablenotemark{b} & {\it b}\tablenotemark{b} & $V_{\rm sys}$\tablenotemark{c}& $\sigma_v$\tablenotemark{d} & $\overline{W}$\tablenotemark{e} & $A$ \tablenotemark{f}&$N({\rm H}_2)$\tablenotemark{g}& $D$\tablenotemark{h}& R$_{\rm gal}$\tablenotemark{i} & $z_{\rm gal}$\tablenotemark{j} & $r$\tablenotemark{k} & $M$\tablenotemark{l}\\
&(\degree)&(\degree)& (\km\ps) &(\km\ps) &(\K\km\ps) &(arcmin$^{2}$) &(\cm$^{-2}$) &(kpc) &(kpc) &(pc) &(pc) &($\Msun$)\\
\tableline
MWISP G122.647+03.350& 122.647& 3.350& $-$59.4& 0.7& 0.99& 19.5& 1.9$\times10^{20}$& 5.4& 12.0& 340& 2.4& 200\\
MWISP G121.752+03.454& 121.752& 3.454& $-$59.7& 0.65& 1.3& 4.5& 2.5$\times10^{20}$& 5.5& 12.0& 360& 0.74& 65\\
MWISP G125.818+00.164& 125.818& 0.164& $-$60.2& 0.42& 0.38& 2.2& 9.2$\times10^{19}$& 5.5& 12.2& 45& 0.6& 12\\
MWISP G126.063+00.229& 126.063& 0.229& $-$59.2& 0.68& 1.5& 6.8& 3.0$\times10^{20}$& 5.4& 12.2& 51& 0.99& 110\\
MWISP G126.916+00.361& 126.916& 0.361& $-$58.3& 0.9& 2.8& 45.0& 5.1$\times10^{20}$& 5.3& 12.1& 63& 2.9& 1200\\
MWISP G125.597+01.150& 125.597& 1.150& $-$59.5& 0.57& 0.87& 2.8& 1.8$\times10^{20}$& 5.5& 12.2& 140& 0.52& 28\\
MWISP G127.395+01.157& 127.395& 1.157& $-$58.6& 0.89& 2.6& 16.8& 4.9$\times10^{20}$& 5.4& 12.2& 140& 1.5& 450\\
MWISP G120.545+01.640& 120.545& 1.640& $-$59.8& 0.39& 0.57& 1.5& 1.2$\times10^{20}$& 5.5& 11.9& 180& 0.42& 10\\
MWISP G120.456+01.662& 120.456& 1.662& $-$59.6& 0.59& 0.96& 7.5& 1.9$\times10^{20}$& 5.4& 11.9& 180& 1.4& 79\\
MWISP G120.133+01.750& 120.133& 1.750& $-$59.4& 0.75& 1.3& 8.5& 2.5$\times10^{20}$& 5.4& 11.8& 190& 1.1& 120\\
...\\
\\
\tableline
\end{tabular}
\tablenotetext{a}{Clouds with incomplete coverage, i.e.\ extending to the boundary of the observation coverage, are marked with an asterisk in their names.}
\tablenotetext{b}{Intensity-weighted mean Galactic longitude and latitude by \twCO~(J=1--0) line.}
\tablenotetext{c}{Intensity-weighted mean LSR velocity of \thCO~(J=1--0) emission; if no significant \thCO~(J=1--0) emission is detected, that of \twCO~(J=1--0) emission is applied instead.}
\tablenotetext{d}{Velocity dispersion, derived as the square root of the intensity-weighted velocity variance (moment II).}
\tablenotetext{e}{Average integrated intensity.}
\tablenotetext{f}{Angular area of the cloud.} 
\tablenotetext{g}{{\rm H}$_2$ column density, estimated using the conversion factor $N({\rm H_2})/W(^{12}{\rm CO})=1.8\E{20} \cm^{-2}(\K \km \s^{-1})^{-1}$ \citep{Dame+2001}.}
\tablenotetext{h}{Kinematic distance estimated based on the systemic velocity, refer to the text in Section~\ref{sec:mcloc} for details.}
\tablenotetext{i}{Distance to the Galactic center, derived as $R_{\rm gal}=\sqrt{D^2 {\rm cos}^2(b)+R_{\rm 0, gal}^2-2 D R_{\rm 0, gal}{\rm cos}(b) {\rm cos}(l)}$, where $R_{\rm 0,~gal}=8.15$~\kpc\ is the Sun's distance to the Galactic center \citep{Reid+2019}.}
\tablenotetext{j}{Height from the Galactic plane, corrected by the Sun's height above the Galactic midplane as 15.7~pc \citep{Zhou+2023, Su+2019}.}
\tablenotetext{k}{Cloud radius, derived as the average of the half widths at half maximum (HWHM) along the projected minor and major axes. The HWHM of the axes are estimated by the intensity-weighted second moment around the center and converted to HWHM by multiplying by 1.1774. The lengths of the axes are deconvolved from the beam to correct for the finite angular resolution, so the corresponding radius is corrected as well. 
For MCs with spatial extensions in the {\it l} or {\it b} directions equal to two pixels (i.e.\ not fully resolved), no radius value is given.}
\tablenotetext{l}{Cloud mass, estimated as $M=N({\rm H_2}) A D^2 \mu({\rm H_2}) m_{\rm H}$, where $\mu({\rm H_2})=2.8$ is the average molecular weight per hydrogen molecule \citep{Kauffmann+2008}, and $m_{\rm H}$ is the hydrogen atom mass.}
(This table is available in its entirety in machine-readable form.)
\end{center}
\end{table*}

MCs identified here are defined as independent continuous structures in the \twCO\ (J=1--0) data cube with $T_{mb} \ge 3\sigma$, following the cloud extraction method introduced by \cite{Yan+2020}. 
Voxels are considered connected when they share a common face.
We fine-tuned several parameters for our observational coverage to extract MCs.
The bottom level of $T_{mb}$ is determined according to the signal-to-noise distribution to ensure reliability (see Section~\ref{sec:obs}). 
To select MCs with enough significance, the following criteria are also applied \citep[see][for a detail examination of the criteria]{Yan+2020}: (1) peak intensity $\ge 5\sigma$; (2) velocity channel number $> 3$; (3) the signal-to-noise ratio of the total cloud emission $\Sigma~T_{mb} > 5\sigma = 5\sqrt{\Sigma~T_{mb, rms}^2}$; (4) spatial extensions in both l and b directions $\ge 2$ pixels.
MCs that are significantly affected by bad channels, i.e.\ with velocities concentrated in the ranges of two channels away from bad channels or with prominent artificial strips along {\it l} or {\it b} directions, are eliminated.
MCs that extend to the boundary of the observation coverage are excluded from subsequent analysis, since their full extent is unclear.
We note that there is a large connected structure spanning the entire G120 region and two structures spanning the entire G50 region, which extend to the boundary of the regions.
Two of these large connected structures contain the curtain MCs discussed in Section~\ref{sec:over}. 
The ability to distinguish different MCs depends on both the spatial and spectral resolutions of the observation. 
In crowded regions (e.g., near the Local arm), physically distinct MCs may be misidentified as a single entity.
Most of such blended MCs are grouped into the large connected structures in the G120 and G50 regions and are excluded from subsequent analysis, as noted above.
Conversely, a physically combined cloud or complex may be identified as multiple different clouds. This separation is considered to be reasonable because the identified clouds are kinematically separated.
\cite{Sun+2021} and \cite{Yan+2022} have compared MWISP data with other surveys (i.e., the CfA 1.2~m CO survey and the FCRAO 14~m outer Galaxy survey) along with their corresponding MC catalogs.
In general, surveys with high sensitivity detect more diffuse emission and more small MCs, while large clouds tend to be resolved into different individual clouds with high angular resolution.
\cite{Yan+2020} has compared the connectivity-based cloud extraction method with the other method, i.e.\ the Spectral Clustering for Interstellar Molecular Emission Segmentation (SCIMES) algorithm \citep{Colombo+2015, Colombo+2019}.
The cloud extraction method applied here conserves the flux well, but it does not incorporate the emission dendrogram. 

\begin{figure*}[ptbh!]
\centerline{{\hfil\hfil
\psfig{figure=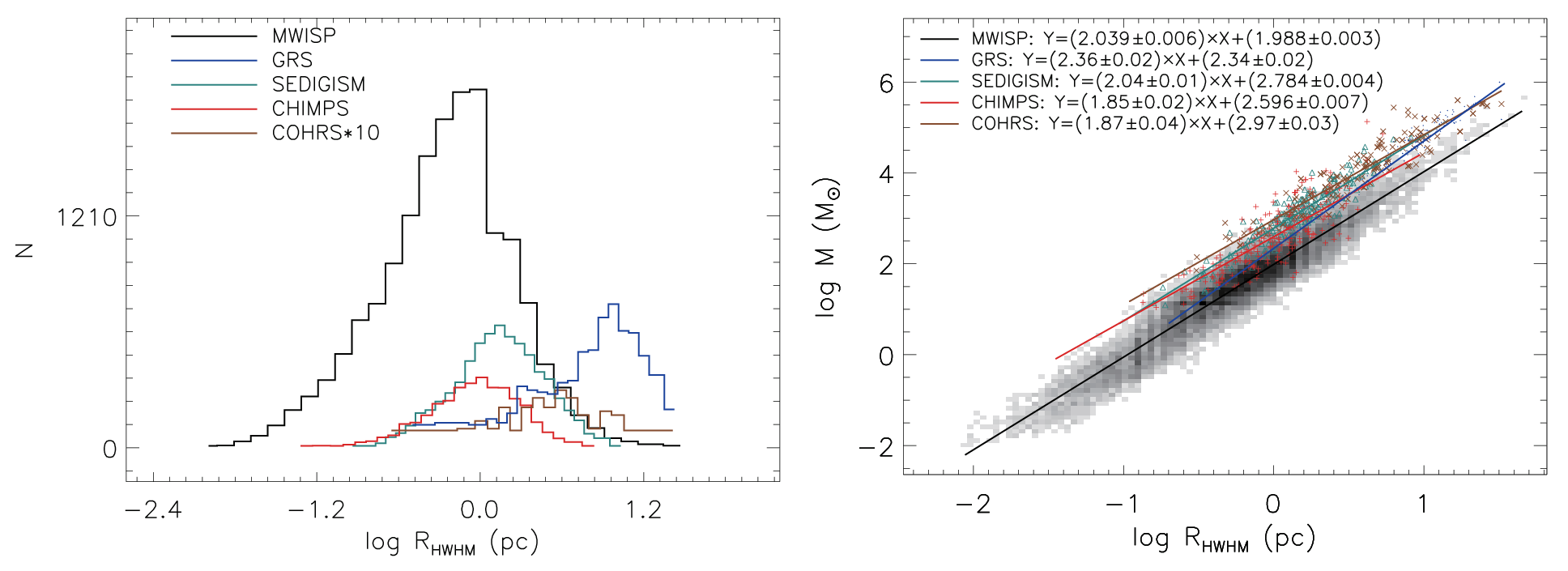,height=2.2in,angle=0, clip=} 
\hfil\hfil}}
\caption{Histogram of radius ({\it left}) and radius-mass relation ({\it right}) for the MWISP clouds, overplotted with those for clouds from other literature catalogues, i.e., GRS \citep[\thCO\ (J=1--0);][]{Rathborne+2009, Roman-Duval+2009, Roman-Duval+2010}, COHRS \citep[\twCO\ (J=3--2);][]{Colombo+2019}, CHIMPS \citep[\thCO\ and \CeiO\ (J=3--2);][]{Rigby+2019}, and SEDIGISM \citep[\thCO\ (J=2--1);][]{Duarte-Cabral+2021}. 
The radius of the clouds are the HWHM, except for the GRS clouds. The radius of the GRS clouds is the equivalent radius estimated by the emission area, which is larger than the HWHM radius.
In the right panel, the density of points for the MWISP sample is shown in a square root scale. The other samples are shown by blue points (GRS), green triangles (SEDIGISM), red plusses (CHIMPS), and brown crosses (COHRS). 
For better visibility, only two hundred randomly selected clouds from each sample are shown.
{\bf Linear fits to the radius-mass relationship are performed on the full sample from each catalogue.}
The fitting results are shown by solid lines, and the corresponding linear functions are indicated in the upper-left corner.}
\label{f:clcomp}
\end{figure*}

The final sample built here for the G120 and G50 regions contains 24724 MCs. In the sample, 276 MCs extend to the boundary of the observation coverage and are excluded from the analysis below. 
Additionally, there are 4171 MCs not fully resolved, i.e.\  with a longitude or latitude span of two pixels.
The basic parameters of the MCs used for this study are listed in Table~\ref{tab:mcpar}. 
We present only a short version of Table~\ref{tab:mcpar} here, and a full version is available at the Science Data Bank at doi:10.57760/sciencedb.21145. 
Figure~\ref{f:clcomp} shows the distribution of the radius and the radius-mass relation of the MCs and compares them with those of the MCs from other literature catalogues.
Using \twCO\ (J=1--0), we detect more diffuse molecular gas, resulting in a smaller average cloud density. Other CO isotopic lines or higher-level CO transition lines are more sensitive to relatively dense gas because they have higher critical densities. However, all the cloud samples identified by different CO lines exhibit a good power-law relation between cloud radius and mass. 
The exponent of the radius-mass relation tends to be lower for cloud samples identified by higher-level CO transition lines, indicating a lower space-filling factor for the corresponding molecular gas. 
Numerous diffuse cloudlets are detected here, benefiting from the high sensitivity. The small MCs detected are nearby, limited by the spatial resolution, e.g., all the MCs with $R_{HWHM}<0.05$~pc are at distances less than 1.5 kpc.
The power-law relationship still holds for these small MCs.
We note that a more detailed study of other measured and derived parameters of the MCs will be presented by Zhou et al.\ (2025, in preparation).

\subsection{MC location} \label{sec:mcloc}
\begin{figure*}[ptbh!]
\centerline{{\hfil\hfil
\psfig{figure=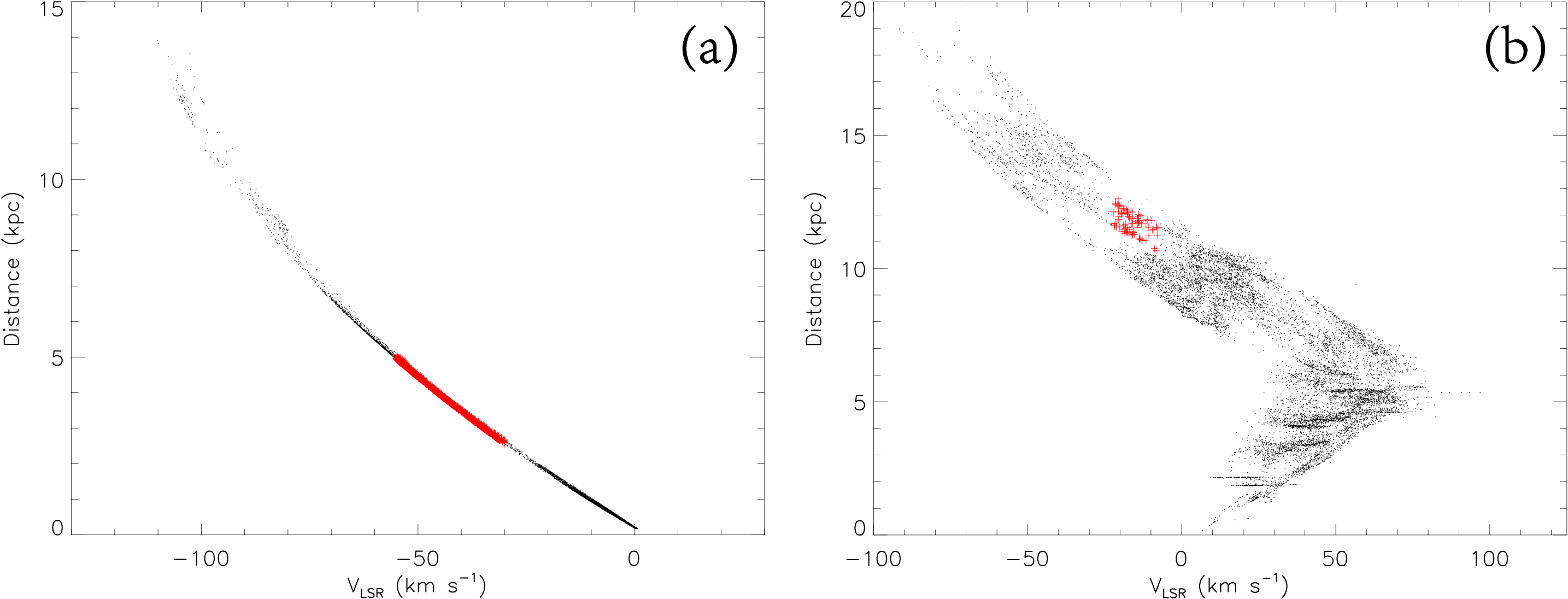,height=2.2in,angle=0, clip=}
\hfil\hfil}}
\caption{Kinematic distance of each identified MC in the G120 ({\it left}) and G50 ({\it right}) regions as a function of systemic velocity. 
The kinematic distances of the MCs are derived from the Galactic rotation curve of \cite{BrandBlitz1993} with the Sun's galactocentric distance of 8.15~kpc and the circular rotation speed of 236 \km\ps\ \citep{Reid+2019}. 
For MCs with positive systemic velocities in the G50 region, based on the rotation curve, one velocity corresponds to two distances, near and far, so we apply  a full distance-probability density function method \citep{Reid+2016, Reid+2019} to calculate the distance instead (see Section~\ref{sec:mcloc} for details).
MCs in the velocity range from $-$55 to $-$30 \km\ps\ in the G120 region are marked as red crosses in the left panel, which are probably located in the Perseus arm spiral shock region.
MCs with velocities ranging from $-$23 to $-$7 \km\ps\ and Galactic longitudes ranging from 48\degree\ to 53.5\degree\ are labeled with red crosses in the right panel, which are in substructures within the inter-arm region between the Perseus and Outer arms, referred to as the G50 Perseus-Outer interarm spurs.}
\label{f:visvd}
\end{figure*}
We derive the kinematic distances of MCs using the Galactic rotation curve of \cite{BrandBlitz1993} with the Sun's galactocentric distance of 8.15~kpc and the circular rotation speed of 236~\km\ps\ \citep{Reid+2019}. 
We note that the kinematic distance of MCs within $\sim$3~kpc has a larger relative error compared to that of more distant MCs.
For MCs with positive systemic velocities in the G50 region, the rotation curve indicates that a single velocity corresponds to two distinct distances: near and far. To address this, we apply a full distance-probability density function method \citep{Reid+2016, Reid+2019} to calculate the distance instead. 
For these MCs, we additionally exclude those with mass exceeding the $1\sigma$ range of the Larson $\sigma_v$-M relation fitted here and those with heights exceeding the $5\sigma$ thickness of the Galactic thick molecular disk \citep[i.e., $z_{\rm gal}\gtrsim$590~pc;][]{Su+2021}, to largely eliminate bias from samples affected by near- and far-distance ambiguities. 
According to this, 4106 MCs are eliminated from the analysis, 50 of which extend to the boundary of the observation coverage.
Since this method determines distance probabilities based on the measured deviation of clouds from both the spiral arms and the Galactic disk, the relevant sample is unsuitable for analysing inter-arm structures or structures distant from the Galactic plane.
Additionally, there are 647 MCs with distances less than the radius of the Local Bubble \citep[i.e.\ 165~pc;][]{Zucker+2022}. These MCs are excluded from the analysis because their velocities are dominated by proper motions rather than the Galactic rotation.
The Perseus arm in the G120 region has a peculiar velocity field that has been well studied in previous work \citep[see][and references therein]{Schwarz+1995}. The observed velocities of different types of objects (e.g., \HII\ regions) are lower than the expected velocities estimated based on their distances using the Galactic rotation curve. \cite{Roberts1972} explained this phenomenon using a spiral shock model. The local standard of rest (LSR) velocity can drop abruptly at the shock front \citep[$\sim2$~kpc;][]{Xu+2006} and slowly approaches the Galactic rotation curve at larger distances \citep{Roberts1972}.
MCs at $\sim-50$~\km\ps\ in the G120 region, located in the Perseus arm, can be influenced by the spiral shock, which may deviate from the Galactic rotation curve and locate at smaller distances. 
MCs around the shock front are in the velocity range from $\sim$$-$55 \km\ps\ to $\sim$$-$30 \km\ps\ \citep{Roberts1972, Schwarz+1995} and are the most affected population. However, not all MCs within this velocity range reside near the shock front; we cannot distinguish them here. Considering the large distance deviations that may exist, these MCs are marked in the spatial distribution analysis below.
The velocity-distance relation is illustrated in Figure~\ref{f:visvd}.

\begin{figure*}[ptbh!]
\centerline{{\hfil\hfil
\psfig{figure=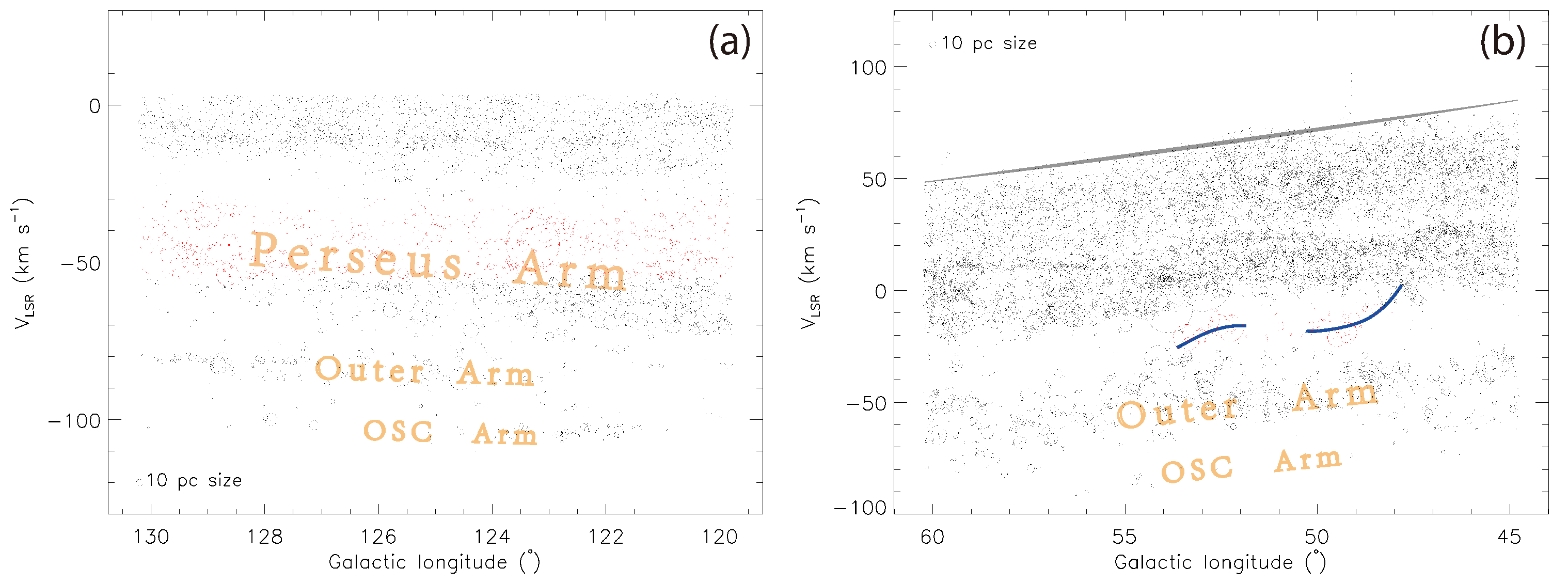,height=2.2in,angle=0, clip=} 
\hfil\hfil}}
\caption{Systemic velocity versus the Galactic longitude for identified MCs in the G120 ({\it left}) and G50 ({\it right}) regions. 
The sizes of the circles indicate the physical sizes of the MCs (see Table~\ref{tab:mcpar}).
MCs that are not fully resolved (i.e., those with a longitude or latitude span of two pixels) are not shown here.
The 10~pc size used as a reference is shown in the lower left corner of the left panel and in the upper left corner of the right panel. 
The red circles mark the same selected MCs as those indicated by the red crosses in Figure~\ref{f:visvd}.
The extensions of the G50 Perseus-Outer interarm spurs are indicated by blue solid lines.
The approximate locations of the spiral arms resemble those in Figures~\ref{f:g120intlv} and \ref{f:g50intlv}. Here, the Outer and OSC arms are labeled and more clearly visible.
The Perseus arm in the G120 region is labeled as well, which has a relatively large velocity span, from approximately $-$75 to $-$30 \km\ps.
The grey solid line is the same as that shown in Figure~\ref{f:g50intlv}, which follows the upper velocity limit well.}
\label{f:visvdcloud}
\end{figure*}

\begin{figure*}[ptbh!]
\centerline{{\hfil\hfil
\psfig{figure=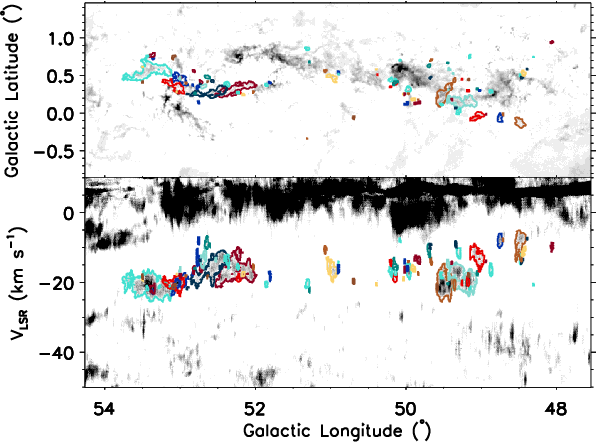,width=5.5in,angle=0, clip=}
\hfil\hfil}}
\caption{Integrated intensity ({\it l-b; top}) and Galactic longitude-velocity ({\it l-v; bottom}) maps of MCs in the G50 Perseus-Outer interarm spurs, with the intensity of all identified MCs shown in greyscale. The {\it l-b} and {\it l-v} maps are of the same field. The selected inter-arm MCs are the same as those indicated by the red crosses in the right panel of Figure~\ref{f:visvd}, of which the masks are overlaid in random colors. The color scheme is the same in the {\it top} and {\it bottom} panels. Background intensities are in linear scale, with the maximum values of 73.5~\K\km\ps\ and 872.3~\K\ in the {\it l-b} and {\it l-v} maps, respectively. Intensities greater than the maxima are truncated for better visibility.}
\label{f:visinterarmcl}
\end{figure*}

As shown in Figure~\ref{f:visvdcloud}, MCs along the distant spiral arms are more clearly visible than in l-v maps (bottom panels of Figures~\ref{f:g120intlv} and \ref{f:g50intlv}), since weak signals are suppressed by the large dynamic range of those images \citep[refer to][for more details of the spiral arms]{Sun+2024}.
In the G120 region, MCs in the Perseus arm are distributed over a velocity range that is apparently broader than that of MCs in the other spiral arms.
In the G50 region, some MCs are aggregated within the inter-arm region between the Perseus and Outer arms (i.e., at $V_{\rm LSR}\sim-15$~\km\ps\ and $l\sim50$\degree; see Figure~\ref{f:visvdcloud}).
Figure~\ref{f:visinterarmcl} shows the longitude-latitude (l-b) and l-v distributions of the selected MCs that are in aggregates away from the spiral arms.
The MCs detected here are only dense parts of the ISM, and there should be more diffuse gas underneath \citep[e.g.,][]{Smith+2014}. 
This aggregation reveals substructures within the inter-arm region, hereafter referred to as the G50 Perseus-Outer interarm spurs.
The spurs appear to extend to the Perseus and Outer arms to the west and east, respectively (see Figure~\ref{f:visvdcloud}). 
Further investigation is needed to examine the continuity of this structure, e.g.\ by \HI\ emission.
The number of the MCs in this substructure is 98, and their total mass is greater than $2.5\E{5}~\Msun$.
In the G50 region, after increasing the velocity of the tangent points by 15~\km\ps, they follow the terminal velocity of MCs well. 
There are fewer MCs exceeding the tangent point velocity at $l\sim45$\degree.
For MCs distributed on such a large scale, their velocities can be affected by spiral arm perturbations, however, normally, the perturbations can only reduce the observed velocity \citep{Ramon-FoxBonnell2018}.
The general excess of the observed terminal velocity over the tangent point velocity indicates that the peculiar velocity of MCs are commonly up to $\sim$15~\km\ps\ in this region, which may originate from random or large-scale streaming motion.
\cite{Gaia2023} measured velocity dispersions of OB stars around the Local arm, obtaining average values of 14.2, 9.4, and 6.2~\km\ps\ in the radial, azimuthal, and vertical directions, respectively.
These OB stars inherit the velocity dispersion of their parent MCs.
The velocity dispersion of MCs probably has axis ratios of the velocity ellipsoid similar to those of massive stars.
Even after increasing the tangent point velocity by 15~\km\ps, there are still multiple MCs that exceed this terminal velocity. 
The streaming motions of these MCs have small scales. Those at $l\sim49$\degree\ are mostly located around the edge of SNR~W51C (see Figure~\ref{f:g50intlv} for the extent of SNR~W51C) and were suggested to be associated with the remnant in previous studies \citep[][and references therein]{Zhou+2023}. Those at $l\sim60$\degree\ are mainly distributed along a filamentary structure near the Galactic plane, which needs a further examination. 

The size of MC is also shown in Figure~\ref{f:visvdcloud}, with more small MCs being detected at closer distances. The MC sample is more complete at closer distances. With an angular resolution of $\sim1'$, MCs larger than 1~pc in size can be detected within a distance of 4~kpc.
The maximum sizes (full widths at half maximum, FWHM) of MCs in the G120 and G50 regions are $\sim78$~pc and $\sim90$~pc, respectively. 
A small number of MCs are detected beyond the Outer arm, and their sizes are relatively small, e.g., a maximum size of $\sim$18~pc.
A comprehensive analysis of MC size distributions will be presented in Zhou et al.\ (2025, in preparation).

\begin{figure*}[ptbh!]
\centerline{{\hfil\hfil
\psfig{figure=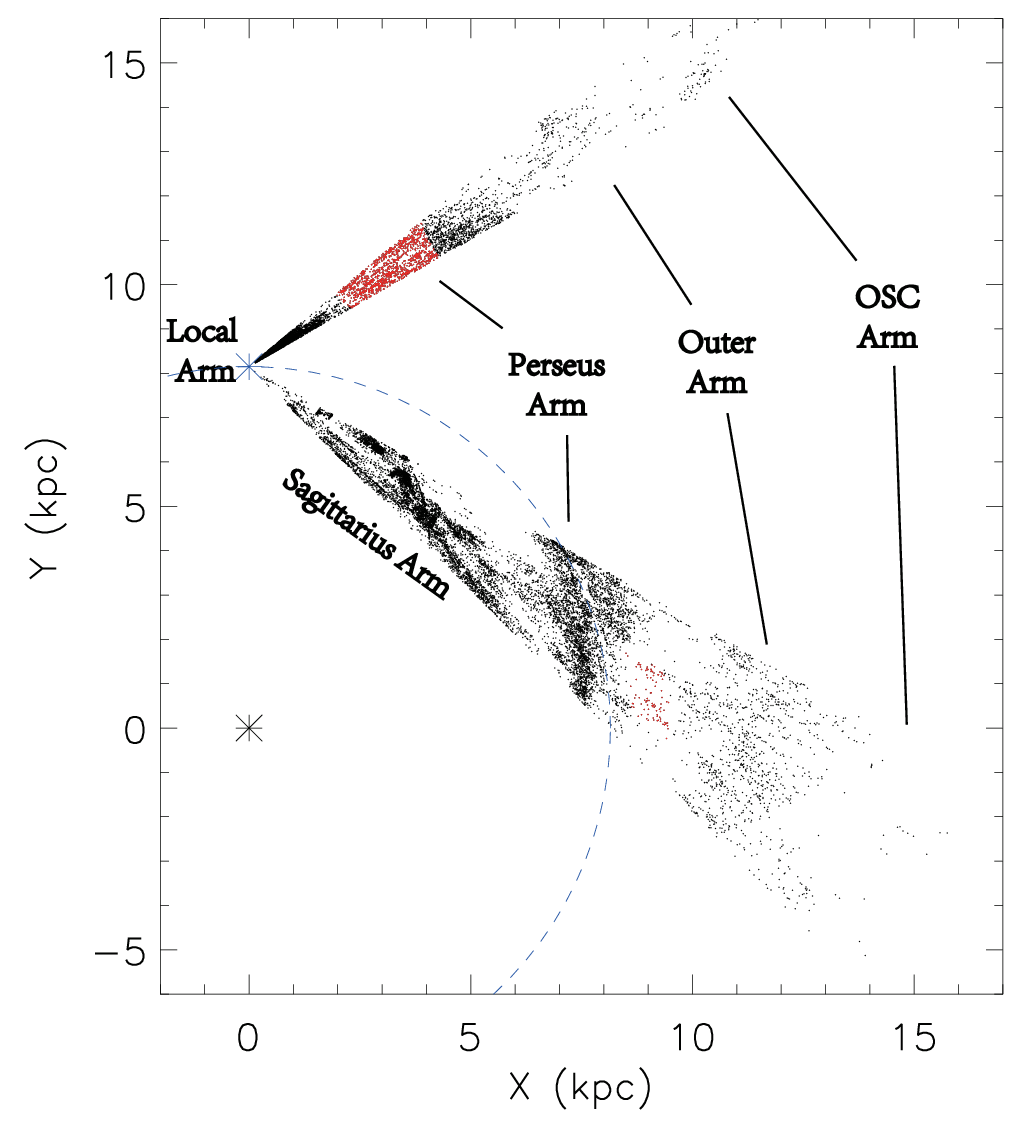,height=5in,angle=0, clip=}
\hfil\hfil}}
\caption{Plan view of MC locations in the G120 and G50 regions from the northern Galactic pole. 
MC distances are estimated using the Galactic rotation curve and the full distance-probability density function method only for MCs with positive systemic velocity in the G50 region.
The red dots mark the same selected MCs as those indicated by the red crosses in Figure~\ref{f:visvd}. The Galactic center and the Sun are marked by black and blue stars, respectively. A blue dashed circle indicates the solar circle. The approximate positions of the spiral arms are shown.}
\label{f:visxy}
\end{figure*}
In the plan view of the location of MCs from the north Galactic pole (Figure~\ref{f:visxy}), the spiral arm structures are clearly visible. 
Additionally, the G50 Perseus-Outer interarm spurs are evident.
In the G120 region, some MCs in the Perseus arm are probably located within a spiral shock region, resulting in a deviation from the Galactic rotation curve and at a distance of $\sim2$~kpc.

\subsection{Galactic disk} \label{sec:disk}
\begin{figure*}[ptbh!]
\centerline{{\hfil\hfil
\psfig{figure=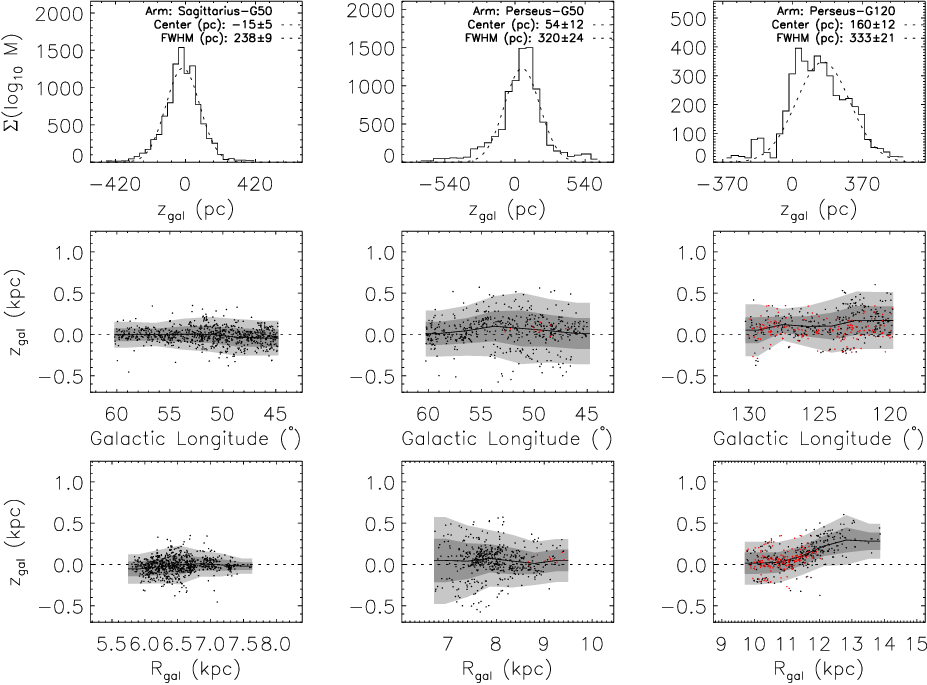,height=4in,angle=0, clip=}
\hfil\hfil}}
\caption{Histogram of height from the Galactic plane for identified MCs ({\it top}), plot of height versus the Galactic longitude ({\it middle}), and plot of height versus galactocentric distance ({\it bottom}) in different segments of the Galactic disk. 
The disk segments are divided at galactocentric distances where the number of MCs reaches a local minimum for the G120 and G50 regions, respectively. 
MCs in the segments are primarily distributed along the known spiral arms, nevertheless, there are also MCs located in the adjacent inter-arm regions.
The segments are named after the spiral arms in them and are labeled in the upper-right corner of the panels.
The Local-G50 and Sagittarius-G50 segments are divided by the heliocentric distance, rather than the galactocentric distance, at the minimum number of MCs.
The height of the Sun above the Galactic midplane is taken as 15.7~pc \citep{Zhou+2023, Su+2019} to fix the position of the Galactic plane. 
The histograms of the heights from the Galactic plane are weighted by the logarithm of mass, by summing the log(M) of each cloud in each bin.
Each bin of the histogram contains at least nine MCs.
The histograms are fitted with Gaussian functions, and the results of the Gaussian fit are shown by black dotted lines. The corresponding Gaussian parameters are indicated in the upper-right corner of the top panels.
In the middle and bottom panels, each spot represents a sample of nine MCs.
Black solid lines indicate the running mean of $z_{gal}$, and the deep and light grey shaded areas indicate the $1\sigma$ and $2\sigma$ ranges, respectively.
The $z_{gal}$=0 plane is shown by the black dotted lines for reference.
The red dots mark the same MCs as those indicated by the red crosses in Figure~\ref{f:visvd}.
}
\label{f:vishgala}
\end{figure*}
\begin{figure*}[ptbh!]
\centerline{{\hfil\hfil
\psfig{figure=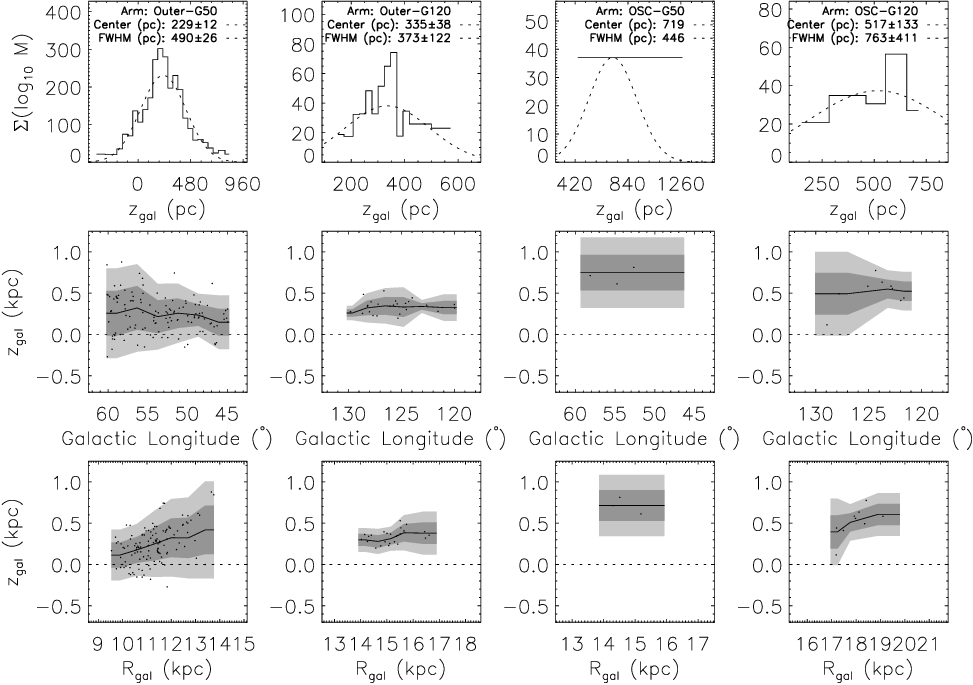,height=4in,angle=0, clip=}
\hfil\hfil}}
\caption{Same as Figure~\ref{f:vishgala}, but for the Outer and OSC arm segments. Due to insufficient samples in the OSC-G50 segment, the center and deviation of its Gaussian function are determined as the mean and deviation of $z_{\rm gal}$, respectively, weighted by the logarithm of mass.}
\label{f:vishgalb}
\end{figure*}

The distribution of MCs perpendicular to the Galactic plane in different segments of the Galactic disk is shown in Figures~\ref{f:vishgala} and \ref{f:vishgalb}. 
The disk segments are divided at galactocentric distances where the number of MCs is at a local minimum, at 9.7, 13.9, and 16.9 kpc for the G120 region, and 6.7, 9.5, and 13.8 kpc for the G50 region. 
The MCs in these segments are primarily distributed along the known spiral arms, nevertheless, there are also MCs located in adjacent inter-arm regions. 
The segments are wider than the spiral arms within them, but do not extend to other arms.
We refer to the segments according to the names of the spiral arms within them below. The Local-G50 and Sagittarius-G50 segments are divided by the heliocentric distance with a minimum number of MCs, 1~kpc, rather than the galactocentric distance.
The histograms of heights from the Galactic plane are weighted by the logarithm of MC mass, by summing the logarithm of mass, to take into account the different contributions of MCs with different masses. 
The logarithm of mass is applied since the distribution of MC mass can be described by power-law functions \citep{HeyerDame2015}. 
To avoid the impact of negative logarithm values, MCs with masses below 1~$\Msun$ are excluded. This only affects the result for the Local arm segments marginally.
Since the observation only covers a small portion of the Galactic disk, the position of the Galactic disk cannot be well determined here. 
The position of the Galactic plane is fixed by establishing the Sun's height above the Galactic midplane as 15.7~pc \citep{Zhou+2023, Su+2019}.
The Galactic latitude boundary of our observation restricts our ability to detect higher MCs at closer distances, hence, the thickness of the segments of the Galactic disk in proximity cannot be well determined, e.g.,\ around the Local arm. 
The thickness of the Galactic molecular disk at a distance can be detected in its entirety.
The sample size for each of the segments is not sufficient to discern between thin and thick Galactic molecular disks \citep[e.g.,][and references therein]{Su+2021}.

As shown in Figure~\ref{f:vishgala}, the height distribution of the MCs in the Sagittarius arm segment is close to a Gaussian distribution. The Sagittarius arm segment is the one closest to the Galactic center, with a galactocentric distance of about 6.5~kpc. It lies basically along the Galactic plane and has a relatively small thickness. 
The Perseus-G50 segment is most affected by the kinematic distance ambiguity problem, the velocity of the MCs in it is close to that in the Local-G50 segment, which contains a large number of MCs. Nevertheless, this has no significant effect on the Gaussian fitting of the whole segment, but it can affect the analysis of distributions on small scales. 
In the Perseus-G120 segment, the peculiar velocity field in the Perseus arm causes the estimated distances of some MCs to be greater than their actual distances, resulting in high $z_{gal}$ values.
The MCs that are possibly affected are marked as red dots in the third column of Figure~\ref{f:vishgala}. This affects the thickness of a section of the segment but does not affect the overall height distribution. The Galactic disk begins to warp at R$_{gal}\sim11$~kpc in this direction.
The MC sample in the Outer arm in the G50 region is relatively complete, with good distance estimations.
The warp of the Galactic disk is significantly enhanced at R$_{gal}\sim10$~kpc in this direction (see the first column in Figure~\ref{f:vishgalb}), and the warp and thickness continue to increase with the galactocentric distance after that. 
The warp and thickness also increase with the Galactic longitude in this segment (see the $l$-$z_{gal}$ plot in the middle-left panel of Figure~\ref{f:vishgalb}). This is probably due to the increasing galactocentric distance along the spiral arm as the Galactic longitude increases.
The height and thickness of the Outer-G120, OSC-G50, and OSC-G120 segments are not well-constrained due to the limited number of MC samples. 

We find that the Galactic molecular disk is thinner in the inter-arm region than in the nearby spiral arm regions (i.e., at $l\sim50$\degree\ and $R_{gal}\sim9$~kpc; see the bottom-middle panel of Figure~\ref{f:vishgala}). 
This inter-arm region is where the G50 Perseus-Outer interarm spurs are located.
We also perform a statistical examination to compare the distributions of the $z_{gal}$ parameter. The variance of $z_{gal}$ of the G50 Perseus-Outer interarm spurs are significantly smaller, which confirms the result. 

\begin{figure*}[ptbh!]
\centerline{{\hfil\hfil
\psfig{figure=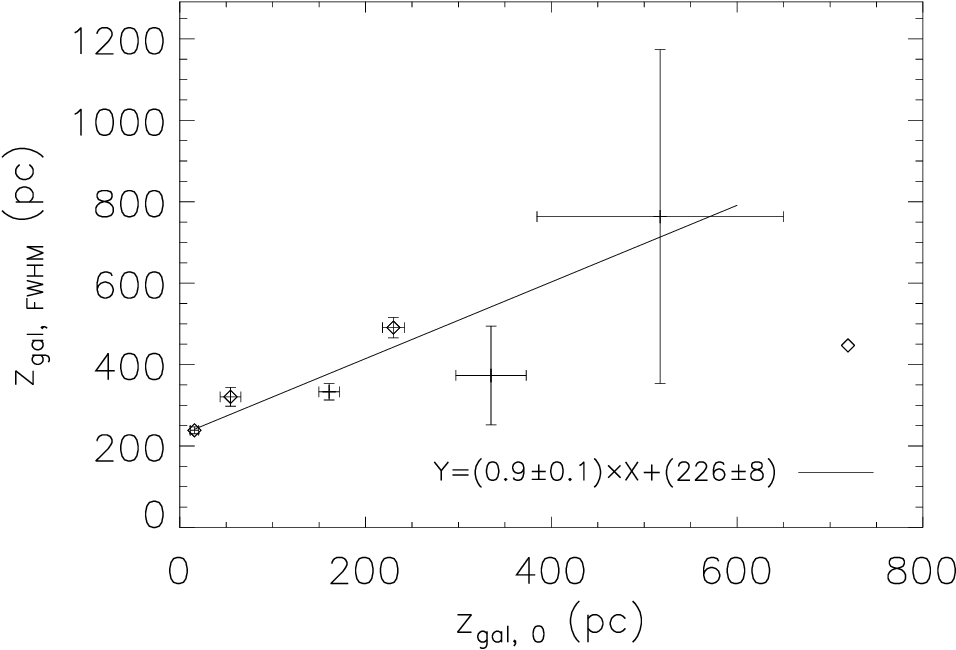,height=3in,angle=0, clip=}
\hfil\hfil}}
\caption{Thickness (FWHM) as a function of height from the Galactic plane for different segments of the Galactic disk in regions G50 (diamonds) and G120 (crosses).
The FWHMs and heights of each Galactic disk segment are obtained by fitting the MC height distribution shown in Figures~\ref{f:vishgala} and \ref{f:vishgalb}. 
The segments around the Local arm, (i.e.\ the Local-G120 and Local-G50 segments) are excluded because their thicknesses are not well determined. 
A linear fit is performed for all segments except the OSC-G50 segment, which has insufficient MC samples to determine height and thickness through Gaussian fitting.
The fitting result is shown by a black solid line, and the corresponding linear function is indicated in the lower-right corner.
}
\label{f:visrh}
\end{figure*}
The warp and flare of the Galactic disk are well-known, both of which are prominent in the outer disk. They have been observed using different tracers, such as stars, ionized gas, \HI\ gas, molecular gas, etc. \citep{Wouterloot+1990, Levine+2006, Chen+2019}. However, there is still no consensus on their formation mechanism.
Some features of the warp and flare detected here are consistent with those observed using \HI\ emission in previous studies. 
Similar to the molecular disk detected here, the warp of \HI\ disk also begins at a galactocentric distance of about 10~kpc. The amplitude of the \HI\ warp at R$_{\rm gal}\sim16$~kpc is also larger in the G50 region than in the G120 region \citep{Levine+2006, Han+2023}.
Additionally, we find that the thickness and warp of the molecular disk correlate well (see Figure~\ref{f:visrh}).
Both the warp and flare are enhanced at larger galactocentric distances. However, the warp or thickness of the disk can differ at the same galactocentric distance (e.g, between the Perseus-G120 and Outer-G50 segments), yet these segments still conform to the established relation.
It indicates that the warp and flare of the Galactic disk have homologous origins \citep[e.g., originating from a tilted dark halo;][]{Han+2023}.
The correlation between the thickness (FWHM) and warp amplitude follows a linear relation with an intercept of $\sim$220~pc at $z_{gal, 0}=0$~pc. This intercept represents the intrinsic thickness of the Galactic molecular disk when unwarped, which is consistent with the thickness of the thick molecular disk or the thin \HI\ disk in the inner Galaxy \citep[e.g.,][]{DickeyLockman1990, LockmanGehman1991, Su+2021}. 
We note that most of the molecular gas mass is in a thin molecular disk with a thickness (FWHM) of $\sim$85~pc, which was studied in previous works \citep[see][and references therein]{HeyerDame2015, Su+2021}.
The OSC-G50 segment exhibits a large vertical displacement and small thickness, however, its limited number of MC samples precludes robust constraints on the height distribution and particularly the thickness parameter.

\begin{figure*}[ptbh!]
\centerline{{\hfil\hfil
\psfig{figure=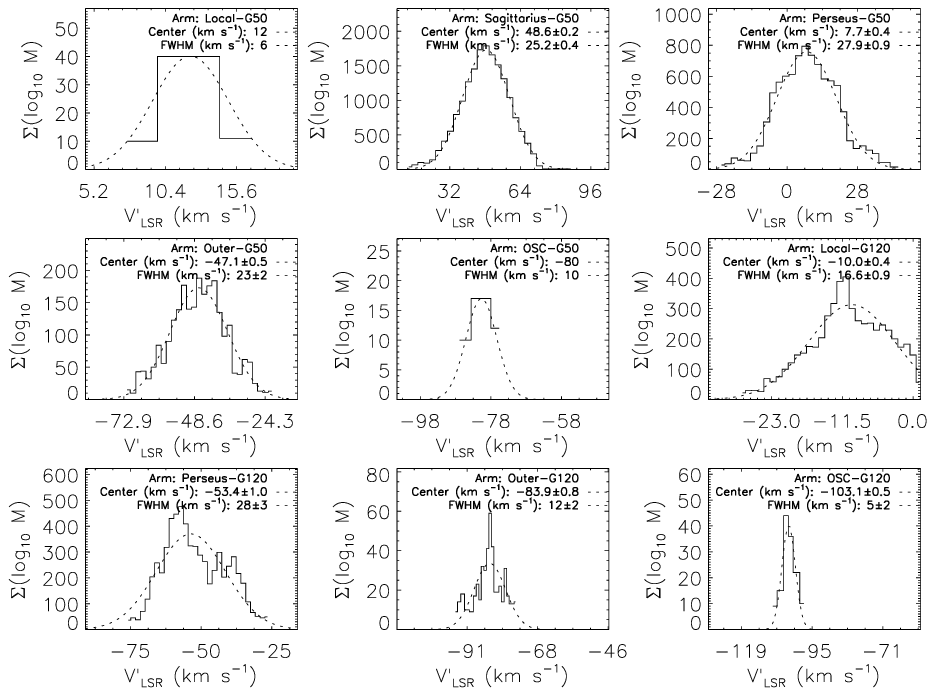,height=4in,angle=0, clip=}
\hfil\hfil}}
\caption{Histogram of modified systemic velocities for MCs in different segments of the Galactic disk. 
The systemic velocity of the MCs is modified by subtracting the baseline of velocity changes with the Galactic longitude for each disk segment, to eliminate the trend of velocity variations along the Galactic longitude.
The baseline is obtained from a linear fit of the velocity variation with the Galactic longitude.
The disk segments are the same as those identified in Figures~\ref{f:vishgala} and \ref{f:vishgalb}, and the names are marked in the upper-right corner of each panel. 
The histograms are weighted by the logarithm of mass, in the same way as in Figures~\ref{f:vishgala} and \ref{f:vishgalb}. 
The histograms are fitted with Gaussian functions, and the results of the Gaussian fit are shown by black dotted lines. 
Due to insufficient samples in the Local-G50 and OSC-G50 segments, the center and deviation of the Gaussian functions are determined as the mean and deviation of the modified velocities, respectively, weighted by the logarithm of mass.
The corresponding Gaussian parameters are indicated in the upper-right corner of each panel.}
\label{f:visvpos}
\end{figure*}
The velocity of MCs can be divided into three components, including circular motion around the Galactic center, streaming motion \citep[e.g.,][]{Ramon-FoxBonnell2018}, and random motion. 
Circular and streaming motions are both affected by projection effects. The projection angle varies with the Galactic longitude and distance.
We modify the systemic velocity of the MCs by subtracting the linear baseline of the velocity change with respect to the Galactic longitude for each disk segment. 
This eliminates the variation in velocity along the LoS caused by changes in the projection angle along longitude.
Figure~\ref{f:visvpos} shows the distribution of the modified systemic velocities of MCs in different segments of the Galactic disk. 
We also evaluate velocity variations along LoS arising from projection effects, according to the changing projection angle with distance for circular motion and the result of the velocity-longitude trend. It contributes less than a quarter of the velocity dispersion, except in the Outer-G50 segment where it accounts for about one-third.
Since we are inside the Local arm, the coverage of observations on the Local arm is too small and biased for such an analysis.
In general, the velocity distribution of MCs is close to a Gaussian distribution (Figure~\ref{f:visvpos}). 
MC samples are incomplete for some segments, nevertheless, a Gaussian fit can complete the velocity distribution to some extent. 
The velocity distribution of MCs in the Perseus-G120 segment shows deviations and splits into two components.
It can indicate the presence of a secondary arm besides the Perseus arm in this region, however, there is no other evidence to support this.
Alternatively, the velocity separation supports the presence of a spiral shock in this region, which has been studied previously \citep[e.g.,][]{Roberts1972, Schwarz+1995}.

\begin{figure*}[ptbh!]
\centerline{{\hfil\hfil
\psfig{figure=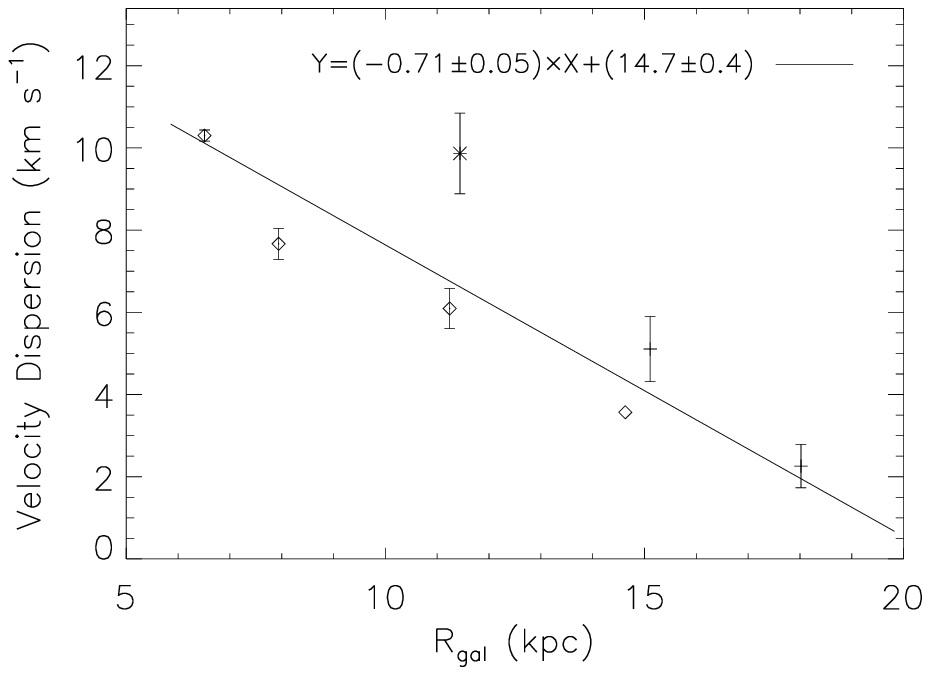,height=3.5in,angle=0, clip=}
\hfil\hfil}}
\caption{Cloud-to-cloud velocity dispersion as a function of galactocentric distance.
The velocity dispersion is estimated by subtracting the distance-dependent velocity variation induced by projection effects from the dispersion of the modified velocity.
The dispersion of the modified velocities for each Galactic disk segment is estimated by Gaussian fitting to the distribution of the modified systemic velocities of the corresponding MCs (FWHM$/2.355$; see Figure~\ref{f:visvpos}). 
The distance-dependent velocity variation induced by projection effects is evaluated using the change in projection angle with distance and the result of the trend of velocity change along the Galactic longitude.
The velocity dispersion of each MC itself is not taken into account.
Diamonds indicate the segments in the G50 region, and crosses indicate the segments in the G120 region. The Perseus-G120 segment is indicated by a star.
The Local arm segments are eliminated due to incomplete coverage. 
A linear fit is performed, and the result of the fit is shown by a black solid line. 
The OSC-G50 segment is excluded from the fit because its velocity dispersion cannot be determined via Gaussian fitting due to an insufficient number of MC samples.
The Perseus-G120 segment is also excluded due to its peculiar velocity field. 
The corresponding fitting parameters are indicated in the upper-right corner.}
\label{f:visrv}
\end{figure*}
After minimizing the dispersion caused by projection effects, we find that the dispersion of MC systemic velocities is correlated with galactocentric distance; as the galactocentric distance increases, the velocity dispersion decreases (see Figure~\ref{f:visrv}).
For the Sagittarius-G50 segment, as indicated by the general excess of observed terminal velocities over the tangent point velocity (Figure~\ref{f:visvdcloud}), the peculiar velocity of MCs in it are commonly up to $\sim$15~\km\ps, which is comparable to the cloud-to-cloud velocity dispersion estimated here. 
At the galactocentric distance of $\sim$8~kpc, \cite{StarkBrand1989} estimated the dispersion of the peculiar velocity of local MCs as $7.8^{+0.6}_{-0.5}$~\km\ps, which is consistent with the velocity dispersion of the Perseus-G50 segment.
Particularly, the Perseus-G120 segment deviates from this relation, which has a larger velocity dispersion.
According to the rotation curve of the Milky Way, the rotation velocity increases slightly with the galactocentric distance in the distance interval we examined \citep[e.g.,][]{BrandBlitz1993, OllingMerrifield2000}.
In addition, the differences in the rotation velocities within each segment is only around 3~\km\ps.
Moreover, the width of a spiral arm has been found to increase with the galactocentric distance \citep{Vallee2020}. The volume of outer disk segments is also larger.
The difference in rotational speed between the inner and outer sides of the segment cannot explain the decreasing velocity dispersion with the galactocentric distance. 
Therefore, the decreasing velocity dispersion indicates reduced random or streaming motion, i.e., less perturbation in the ISM at larger galactocentric distance.
The varying velocity dispersion may be caused by different energy inputs via supernovae at different galactocentric radii \citep{NarayanJog2002}.
The estimated radial gradient of $-0.71\pm0.05$~\km\ps\kpc$^{-1}$ for the velocity dispersion of MCs is consistent with the corresponding \HI\ gradient $-$0.8~\km\ps\kpc$^{-1}$. The latter was derived by modeling the gravitational potential of the gas and stellar components while applying a velocity dispersion of 8~\km\ps\ at R$_{\rm gal}=8.5$~kpc \citep{NarayanJog2002}. 
Molecular gas probably inherits its turbulence from atomic gas, and their consistent velocity dispersions suggest that there is no discontinuity in turbulence properties between them. Furthermore, it has been found that the cloud-to-cloud velocity dispersion decreases as the MCs become more massive \citep[e.g.,][]{Stark1984}.
\cite{YuanYang2025} performed a detailed study of the cloud-to-cloud velocity dispersion in MCs of different sizes.
We note that it is needed to analyze the correlation between the velocity dispersion and galactocentric distance for the MC sample in Galactic longitude ranges outside the G120 and G50 regions to establish a more general relationship.
The velocity dispersion due to random or streaming motion can cause kinematic distance errors. According to the relationship between the velocity dispersion and the galactocentric distance, MCs at larger galactocentric distances have smaller kinematic distance errors.

\section{Summary} \label{sec:sum}
We use unbiased survey data of \twCO, \thCO, and \CeiO\ (J=1–0) line emission to study molecular gas distribution in the Galactic plane.
The observations focus on two regions toward the inner and outer Galaxy, namely the G50 ($l=44$.75\degree--60.25\degree) and G120 ($l=119$.75\degree--130.25\degree) regions, both with the Galactic latitude $|b|\le5.25$\degree, which are part of the MWISP survey. 

In the large-scale distribution, the aggregation of molecular gas along the Galactic plane is noticeable, with some shell-like and filamentary structures widely distributed. In the Galactic plane, MCs tend to concentrate on coherent structures, sketching spiral arm structures.
Molecular gas is also detected at high Galactic latitudes, some of which is known to be nearby.
Particularly, in our proximity, where weak molecular line emission can be detected, several molecular structures with large angular scales and small velocity dispersions are found. Similar MCs have also been detected toward the central region of the Milky Way by the SEDIGISM survey \citep{Schuller+2021}. These large MCs appear to be thin in the LoS direction and widely distributed around us, resembling curtains of mist, and are referred to as curtain MCs here.
Additionally, there are many line broadening features on small scales, which could be driven by star-formation activity or SNRs, etc. These structures suggest a complex and dynamic environment in the ISM.
In the G50 region toward the inner Galaxy, the MC terminal velocities generally exceed the tangent point velocity by $\sim$15~\km\ps. 
The general excess of MC terminal velocities suggest that the peculiar velocity of MCs can be up to $\sim$15~\km\ps\ in this region.
There are some MCs with systemic velocities that exceed the tangent point velocity, even after increased by 15~\km\ps. Some of these MCs are probably associated with SNRs, such as those around SNR~W51C, while the origin of the rest remains uncertain and requires further investigation.

Individual MCs are identified based on the connectivity of the data in the position-position-velocity space. 
The distances of the MCs are estimated, and their distribution across the Milky Way is investigated.
The MC sample built here for the G120 and G50 regions contains 24724 MCs. In the sample, 4171 MCs are not fully resolved, i.e.\ with a longitude or latitude span of two pixels. 
There are also 4753 MCs with unreliable distance measurements, and 276 MCs extending to the boundary of the observation coverage, all of which are excluded from the overall analysis.
There are numerous cloudlets detected, which still follow the overall power-law radius-mass relation.
Sub-arm structures are found in the inter-arm region between the Perseus and Outer arms at $l\sim50$\degree\ and $V_{LSR}\sim-15$~\km\ps, referred to as the G50 Perseus-Outer interarm spurs. 
These structures suggest that MCs can aggregate in the inter-arm region, indicating a distribution that is more complex than a simple spiral pattern.
The Galactic molecular disk in this inter-arm region is thinner than that in the nearby spiral arm region.
The overall distribution of MCs perpendicular to the Galactic disk is examined.
The warp of the Galactic disk begins at R$_{\rm gal}$ of $\sim$11\kpc\ and $\sim$10\kpc\ in the G120 and G50 directions, respectively.
Both the warp and flare are enhanced at larger galactocentric distances. However, the warp or thickness of the disk can differ at the same galactocentric distance.
Remarkably, we find that the thickness and warp of the disk segments are correlated with each other, indicating homologous origins of the warp and flare of the Galactic disk.
The molecular disk has a typical thickness of $\sim$220~pc in the inner Galaxy.
We also find that the dispersion of the MC systemic velocity decreases with increasing galactocentric distance, probably due to reduced perturbation in the ISM at large galactocentric distances. Consequently, the error in the kinematic distance of MCs is smaller at larger galactocentric distances.
The Perseus arm segment in the G120 region is an exception to the relationship between velocity dispersion and galactocentric distance, as it has a large velocity dispersion. Moreover, its velocity distribution splits into two components. 
This can be explained by the presence of a spiral shock there.

\begin{acknowledgments}
We thank the anonymous referee for helpful comments that improved this paper and its conclusions.
This research made use of the data from the Milky Way Imaging Scroll Painting (MWISP) project, which is a multi-line survey in 12CO/13CO/C18O along the northern Galactic plane with PMO-13.7m telescope. We are grateful to all the members of the MWISP working group, particularly the staff members at PMO-13.7m telescope, for their long-term support. MWISP was sponsored by National Key R\&D Program of China with grants 2023YFA1608000 \& 2017YFA0402701 and by CAS Key Research Program of Frontier Sciences with grant QYZDJ-SSW-SLH047.
\end{acknowledgments}
 
\bibliographystyle{aasjournal}
\bibliography{ms.bbl}


\end{document}